\documentclass[aps,prc,reprint,superscriptaddress]{revtex4-1}
\usepackage{dcolumn}
\usepackage{amsmath,amssymb,amsthm,paralist}
\usepackage{graphicx}
\usepackage{appendix}
\usepackage{subfigure}
\usepackage{longtable}

\usepackage{amsmath}
\usepackage{amsfonts}
\usepackage{hyperref}
\usepackage{url}
\usepackage{verbatim}

\begin{document}


\title{$s$-wave scattering lengths for the $^7$Be+p system from an \textit{R}-matrix analysis}

\author{S.~N.~Paneru}
\email[]{sp266413@ohio.edu}

\affiliation{Department of Physics and Astronomy, Ohio University,
Athens, Ohio 45701, USA}
\author{C.~R.~Brune}
\affiliation{Department of Physics and Astronomy, Ohio University,
Athens, Ohio 45701, USA}
\author{R.~Giri}
\affiliation{Department of Physics and Astronomy, Ohio University,
Athens, Ohio 45701, USA}
\author{R.~J.~Livesay}
\affiliation{Department of Physics, Colorado School of Mines,
Golden, Colorado 80401, USA}
\author{U.~Greife}
\affiliation{Department of Physics, Colorado School of Mines,
Golden, Colorado 80401, USA}
\author{J.~C.~Blackmon}
\affiliation{Department of Physics and Astronomy, Louisiana State University,
Baton Rouge, Louisiana 70803, USA}
\author{D.~W.~Bardayan}
\affiliation{Department of Physics, University of
Notre Dame, Notre Dame, Indiana 46556, USA}
\author{K.~A.~Chipps}
\affiliation{Physics Division, Oak Ridge National Laboratory, Oak Ridge, Tennessee 37831, USA}
\author{B.~Davids}
\affiliation{TRIUMF, Vancouver, British Columbia V6T 2A3, Canada}
\affiliation{Physics Department, Simon Fraser University, Burnaby, British Columbia V5A1S6, Canada}
\author{D.~S.~Connolly}
\affiliation{TRIUMF,
Vancouver, British Columbia V6T 2A3, Canada}
\author{K.~Y.~Chae}
\affiliation{Department of Physics, Sungkyunkwan University, Suwon 16419, Korea}
\author{A.~E.~Champagne}
\affiliation{Department of Physics and Astronomy, University of North Carolina, Chapel Hill, North Carolina, 27599, USA}
\author{C.~Deibel}
\affiliation{Wright Nuclear Structure Laboratory, Yale University, New Haven, Connecticut 06520, USA}
\author{K.~L.~Jones}
\affiliation{Department of Physics and Astronomy, University of Tennessee, Knoxville, Tennessee 37996, USA}
\affiliation{Department of Physics and Astronomy, Rutgers University, Piscataway, New Jersey 08854, USA}
\author{M.~S.~Johnson}
\affiliation{Department of Physics and Astronomy, Rutgers University, Piscataway, New Jersey 08854, USA}
\author{R.~L.~Kozub}
\affiliation{Physics Department, Tennessee Technological University, Cookeville, Tennessee 38505, USA}
\author{Z.~Ma}
\affiliation{Department of Physics and Astronomy, University of Tennessee, Knoxville, Tennessee 37996, USA}
\author{C.~D.~Nesaraja}
\affiliation{Department of Physics and Astronomy, University of Tennessee, Knoxville, Tennessee 37996, USA}
\affiliation{Physics Division, Oak Ridge National Laboratory, Oak Ridge, Tennessee 37831, USA}
\author{S.~D.~Pain}
\affiliation{Department of Physics and Astronomy, Rutgers University, Piscataway, New Jersey 08854, USA}
\author{F.~Sarazin}
\affiliation{Department of Physics, Colorado School of Mines,
Golden, Colorado 80401, USA}
\author{J.~F.~Shriner Jr.}
\affiliation{Physics Department, Tennessee Technological University, Cookeville, Tennessee 38505, USA}
\author{D.~W.~Stracener}
\affiliation{Physics Division, Oak Ridge National Laboratory, Oak Ridge, Tennessee 37831, USA}
\author{M.~S.~Smith}
\affiliation{Physics Division, Oak Ridge National Laboratory, Oak Ridge, Tennessee 37831, USA}
\author{J.~S.~Thomas}
\affiliation{Department of Physics and Astronomy, Rutgers University, Piscataway, New Jersey 08854, USA}
\author{D.~W.~Visser}
\affiliation{Department of Physics and Astronomy, University of North Carolina, Chapel Hill, North Carolina, 27599, USA}
\author{C.~Wrede}
\affiliation{Wright Nuclear Structure Laboratory, Yale University, New Haven, Connecticut 06520, USA}


\begin{abstract}

The astrophysical $S$-factor for the radiative proton capture reaction on $^7$Be ($S_{17}$) at low energies is affected by the $s$-wave scattering lengths. We report the measurement of elastic and inelastic scattering cross sections for the $^7$Be+p system in the center-of-mass energy range 0.474- 2.740~MeV and center-of-mass angular range 70$^\circ$- 150$^\circ$. A radioactive $^7$Be beam produced at Oak Ridge National Laboratory's (ORNL)
Holifield Radioactive Ion Beam Facility was accelerated and bombarded a thin polypropylene (CH$_{2}$)$_\text n$ target. Scattered ions were 
detected in the segmented Silicon Detector Array. Using an \textit{R}-matrix analysis of ORNL and Louvain-la-Neuve cross-section data, the $s$-wave scattering lengths for channel spins
1 and 2 were determined to be 17.34$^{+1.11}_{-1.33}$ and -3.18$^{+0.55}_{-0.50}$~fm, respectively. The uncertainty in the $s$-wave scattering lengths reported in this work is smaller by a factor of 5-8  
compared to the previous measurement, which may reduce the overall uncertainty in $S_{17}$ at zero energy. The level structure of $^8$B is discussed based
upon the results from this work. Evidence for the existence of 0$^+$ and 2$^+$ levels in $^8$B at 1.9 and 2.21~MeV, respectively, is observed. 
\end{abstract}
\pacs{}

\maketitle

\section{INTRODUCTION}
\label{Introduction}
The total terrestrial flux of high-energy neutrinos resulting from the $\beta^+$ decay of $^8$B in the Sun has been measured with a precision of $\pm$4$\%$~\cite{SNO2013,SuperK}. 
Comparisons of the measured and predicted $^8$B solar neutrino fluxes are therefore limited primarily by the theoretical 
uncertainty of approximately $\pm$14$\%$ associated with standard solar model predictions~\cite{Haxton}.
The low-energy astrophysical $S$ factor for the $^7$Be(p,$\gamma$)$^8$B radiative capture reaction, $S_{17}(E)$, is the most uncertain nuclear input needed
to predict the $^8$B solar neutrino flux~\cite{Bahcall_1969,adelberger11} in the standard solar model. It must be known at or near the Gamow peak of $\sim$18~keV, which is 
experimentally inaccessible due to the Coulomb barrier \cite{brune15}. The cross sections are unmeasurably small at these energies, so available data starting 
around 100~keV above the Gamow peak must be extrapolated to solar energies with the aid of theoretical models.\par
Descouvemont~\cite{Descouvemont_2004} used a microscopic three-cluster model and a potential model to study the theoretical uncertainty in extrapolating $S_{17}$ to 
zero energy and found that below 1~MeV it is dominated by the uncertainties in the $s$-wave scattering lengths for the
$^7$Be + p system. A leading-order calculation of $^7$Be(p,$\gamma$)$^8$B in a low-energy effective field theory \cite{Zhang_2014} found 
that the experimental uncertainties in the scattering lengths strongly affected the calculation at energies as low as 400~keV. A simple potential model~\cite{Baye_2000}
shows the importance of the $s$-wave scattering lengths in extrapolating $S_{17}$ to zero energy, although it is not clear how the results in this paper can be translated into uncertainties in the $S_{17}$(0) value deduced from capture data. Although one recent effective field theory calculation~\cite{Zhang2015} suggests that the contribution of scattering length uncertainties to the extrapolation uncertainty of $S_{17}$ below 500~keV may not be large, this sensitivity depends on the range of scattering lengths considered in the calculation. \par

Owing to the required use of radioactive $^7$Be (half-life = 53.2 days), the  scattering lengths have only been measured once, 
by Angulo \textit{et al}.~\cite{ANGULO2003211}, who found $a_{01}$= 25$\pm$9 fm and $a_{02}$= -7$\pm$3 fm, where $a_{0I}$ is the $s$-wave scattering length for channel spin $I$. The $s$-wave scattering lengths deduced from the \textit{ab-initio} no-core shell model/resonating group method~\cite{NAVRATIL2011} are $a_{01}$= -5.2~fm and $a_{02}$= -15.3~fm. Discrepancies in the predicted and measured $s$-wave scattering lengths, particularly for channel spin 2, demand caution when using theoretical models in the extrapolation of $S_{17}$  to zero energy.
Reference~\cite{NAVRATIL2011} also calculates the astrophysical $S$ factor for $^7$Be(p,$\gamma$)$^8$B radiative capture reaction at  zero energy, but the relationship between the $s$-wave scattering lengths and $S_{17}$(0) is not highlighted.
 Better constraints on the scattering lengths may lead to a 
significant reduction in the uncertainty of $S_{17}$(0), thereby reducing the overall uncertainty in the $^8$B neutrino flux prediction.\par
The evaluation of $S_{17}$ in the energy range below 100~keV depends on complete knowledge of the low-lying energy levels of $^8$B, which 
remains elusive \cite{tilley04}. 
There have been several $^7$Be+p elastic scattering measurements aimed at elucidating the level structure of $^8$B.
Gol'dberg \textit{et al}.~\cite{ Goldberg1998} measured the elastic scattering excitation function with a thick target at relative kinetic energies $E$ 
from 1 to 3.6~MeV at 0$^\circ$ in inverse kinematics and proposed the existence of a $1^+$ level at $E_x = $ 2.83~MeV with a width of 780~keV. 
Rogachev \textit{et al}.~\cite{Rogachev_2001} measured elastic scattering using a thick target  over a relative kinetic energy range from 1 to 3.3~MeV 
and found evidence for the existence of a $2^-$ level at $E_x = $ 3.5$\pm$0.5~MeV with a width of $8\pm$4~MeV. Angulo \textit{et al}.~\cite{ANGULO2003211} measured 
the $^7$Be+p elastic cross section with a thin polyethylene target from $E$ = 0.3~MeV to $E$ = 0.75~MeV. From an \textit{R}-matrix analysis, the scattering lengths were inferred 
and the width of the $1^+$ resonance at $E = $ 634$\pm$5~keV was determined to be 31$\pm$4~keV. Yamaguchi \textit{et al}.~\cite{Yamaguchi2009} measured resonant 
elastic and inelastic scattering from $E = $ 1.3 to 6.7~MeV, adducing evidence for $2^-$ and $1^-$ states. Based on an \textit{R}-matrix analysis of a recent thick-target 
elastic and inelastic scattering measurement, Mitchell \textit{et al}.~\cite{Mitchell_2013} proposed new low-lying $0^+$, $2^+$, and $1^+$states
at $E_x = $ 1.9, 2.54, and 3.3~MeV, respectively, in $^8$B. These levels have not yet been confirmed by further experiments. Thus far there has been only a single measurement
of elastic scattering below $E =$ 1~MeV and the available data at higher energies are inconsistent. Based on these experiments there are only two well-known
excited states of $^8$B, the $1^+$ and $3^+$ states at 0.77 and 2.32~MeV, respectively. All other states inferred on the basis of previous $^7$Be+p elastic
scattering measurements require further experimental verification.\par
This paper describes a new measurement of the elastic and inelastic scattering cross sections of $^7$Be+p and a determination of the $s$-wave scattering 
lengths using an \textit{R}-matrix analysis. It also presents evidence for the existence of various excited states in $^8$B that must be properly described in 
theoretical models of its structure. The measurement of elastic and inelastic scattering was performed in inverse kinematics from $E_\text{c.m.}$= 0.474~MeV to 
2.740~MeV covering a center-of-mass angular range of 70$^\circ$ to 150$^\circ$. We used the \textit{R}-matrix method~\cite{Lane_Thomas} 
to analyze elastic and inelastic scattering data. 
In this work, we confirm the existence of some of the levels reported in the literature and re-assess that of others. In particular, we find no evidence in our data set for the 1$^+$ level at 3.3 MeV that has been reported in Ref.~\cite{Mitchell_2013}.

\par
The experimental method used to measure the elastic and inelastic scattering is explained in Sec.~\ref{experiment}. We used a multichannel, 
multilevel \textit{R}-matrix approach to analyze elastic and inelastic scattering data simultaneously. The best-fit parameters from the \textit{R}-matrix
analysis were used to determine the $s$-wave scattering lengths using the method described in Sec.~\ref{R-matrix}. Section~\ref{results} contains
the findings of this work and a comparison with available data from the literature. We conclude in Sec.~\ref{conclusion}.

\section{EXPERIMENT}
\label{experiment}
The elastic and inelastic $^7$Be+p scattering cross sections were measured in inverse kinematics between 0.474 and 2.740~MeV in the center-of-mass system at the 
Holifield Radioactive Ion Beam Facility (HRIBF)~\cite{HRIBF_2011} of Oak Ridge National Laboratory (ORNL). $^7$Be was produced at the 
Triangle University Nuclear Laboratory using the $^7$Li(p,n)$^7$Be reaction~\cite{Fitzgerald}. The lithium targets (disks of 2-cm diameter and 3-mm thickness) were bombarded 
with 8- to 11-MeV protons, typically producing 240 mCi of $^7$Be. The activity was transported to ORNL in the form of an ingot for chemical extraction and 
concentration using the method described in Ref.~\cite{chemical_extraction}. $^7$Be ions were injected into the HRIBF's tandem accelerator via a cesium sputter source. The beam was stripped to the 4$^+$ charge state before the analyzing magnet, removing any $^7$Li whose maximum charge state is 3$^+$. 
The fully-stripped $^7$Be beam was then directed into the target chamber hosting the Silicon Detector Array (SIDAR)~\cite{SIDAR_2000}. 
Additional details of the experimental setup are provided in Ref.~\cite{Livesay}.
 \par
\begin{figure}
\begin{center}
\includegraphics[width=1.0\columnwidth,angle=0]{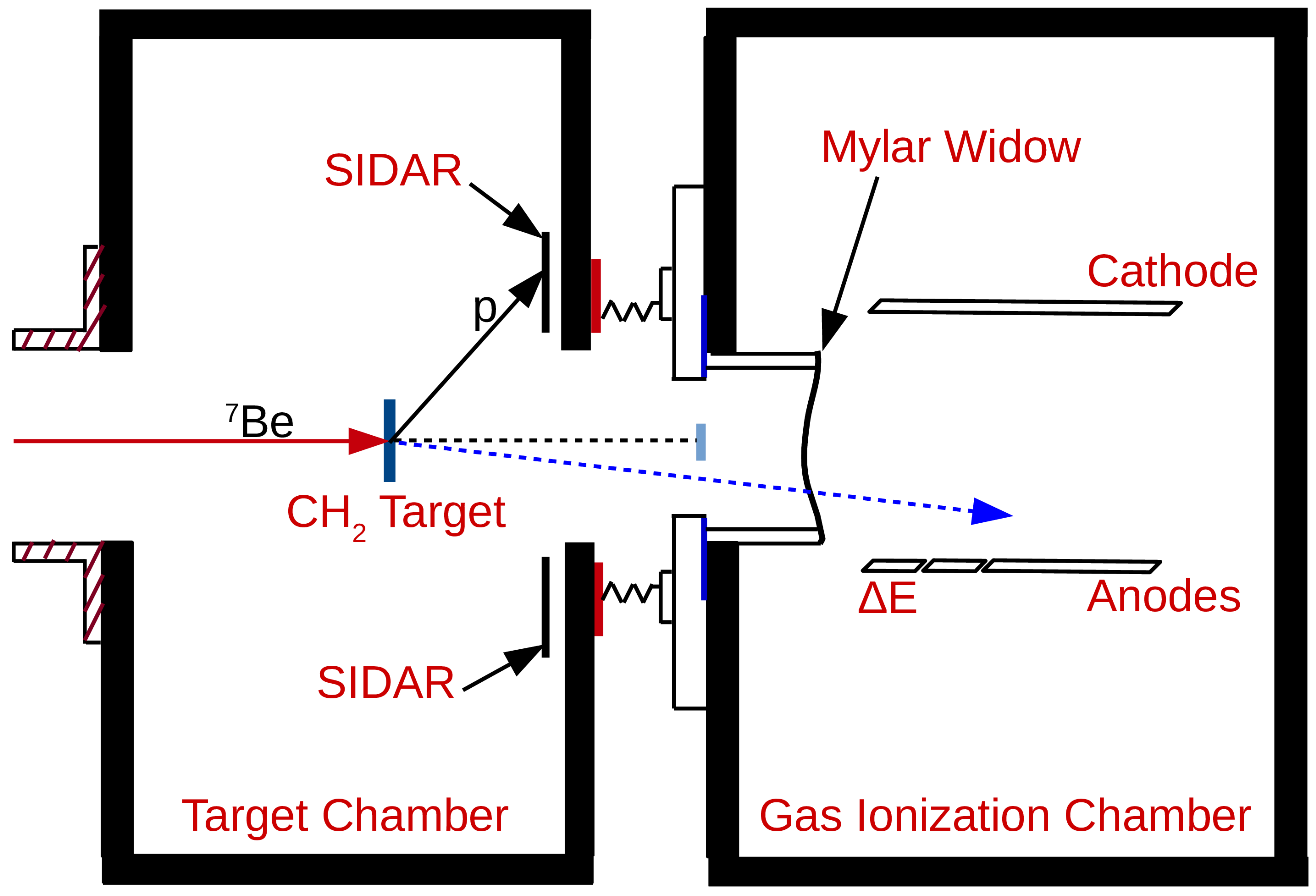}
\caption{ Experimental setup~\cite{SIDAR_2000}. The $^7$Be beam delivered by HRIBF (on left) bombarded a thin 
polypropylene (CH$_2$)$_\text n$ target. Protons were detected by a SIDAR, which was mounted on the downstream face of the scattering chamber. The ion 
chamber placed downstream of the scattering chamber was used for tuning and beam diagnostics.
\label{experimental_setup}}
\end{center}
\end{figure}
The SIDAR consists of an array of Micron YY1 detectors with 40-keV energy resolution, which can be arranged in either a lamp-shade (with six wedges) or a flat 
configuration (with eight wedges). We utilized the SIDAR in the flat configuration for this experiment. The array was composed of detectors of either 300- or 500-$\mu$m 
nominal thickness. A schematic diagram of the target station is shown in Fig.~\ref{experimental_setup}. Self-supporting thin foils of 
polypropylene (CH$_2$)$_\text n$ and gold (Au) were used as the targets. The thickness of the (CH$_2$)$_\text n$ target was determined via $\alpha$-particle energy 
loss measurements to be 100 $\mu$g/cm$^2$, with an uncertainty of $\pm$10$\%$ resulting from the stopping power
calculations. The target foils were mounted
on a retractable target ladder placed in the scattering chamber. There were two diagnostic tools on the ladder, namely, an aperture and a phosphor screen, 
which provide information about the location and size of the beam in the scattering chamber. The scattered protons were detected in the SIDAR 
located downstream of the target. The ionization chamber was separated from the target chamber by a 0.9-$\mu$m-thick mylar window and filled 
with 40 T of isobutane gas. The ionization chamber was used for tuning and beam diagnostics. The unscattered beam was blocked by a 1.5-cm aluminum disk that was
small enough to let the scattered $^7$Be ions enter into the ionization chamber.
\par
The experiment was performed in two campaigns, for which the experimental configurations were similar. The measurements were taken using two different distances of the
SIDAR from the target, providing overlapping angular ranges of $\theta_\text{lab} = $ 26$^{\circ}$-50$^{\circ}$ and $\theta_\text{lab} = $ 14$^{\circ}$-31$^{\circ}$. 
The $^7$Be bombarding energies were chosen in 16 energy steps between 4 and 27~MeV with intensities of 10$^6$-10$^7$ pps at the target station. The $^7$Be+p scattering
cross sections were measured relative to the $^7$Be+Au and $^7$Be+$^{12}$C scattering cross sections, which were used for normalization of the data. 
The energy loss in the target was taken into account by calculating the effective beam energy as $E_\text{eff} = E_{0}-\Delta E/2$, where  $E_{0}$ is the incident beam energy
and $\Delta E$ is the energy loss in the 
target calculated using {\sc SRIM}~\cite{SRIM}. This procedure is valid as long as there is no strong energy dependence of the cross section over the energy
range covered in the target. Since there is a resonance 
at $E_\text{c.m.} = $ 0.634~MeV, the correction factors for the low energy experimental data points were calculated using Eqs.~(6) and (7) from
Ref.~\cite{BRUNE2013}. The correction factor calculated for the 5.2-MeV measurement in the laboratory system was 0.90, while for all other experimental data points, the correction 
factor was within 2$\%$ of unity. This correction factor has been included in the analysis of the $E_{\text{lab}}$=5.2~MeV measurement. \par
\begin{figure}
\begin{center}
\includegraphics[width=1.0\columnwidth,angle=0]{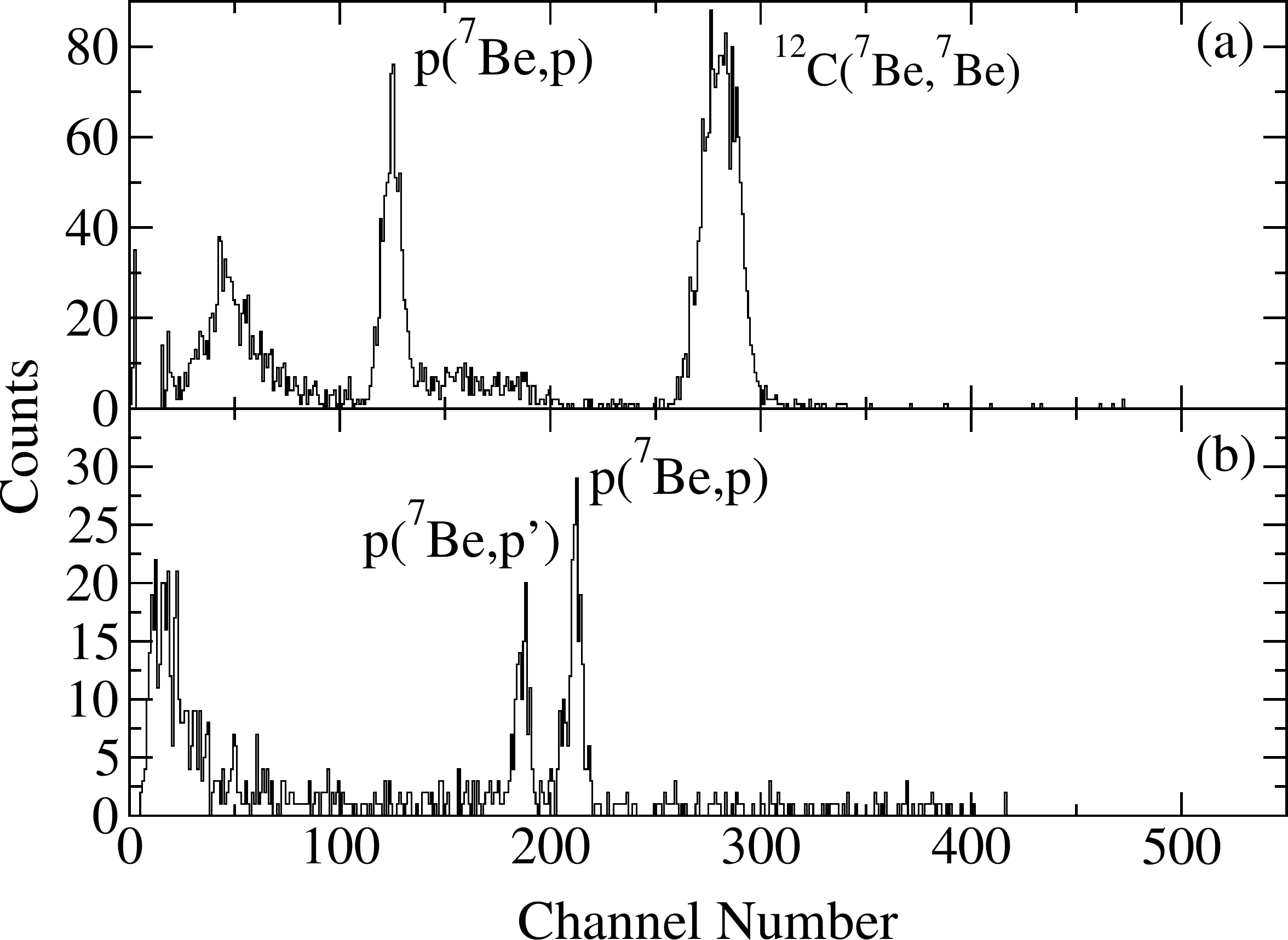}
\caption{Two spectra from the experiment. (a) Spectrum obtained with a (CH$_2$)$_\text n$ target at a $^7$Be beam energy of 5.2~MeV and $\theta_\text{lab} = $ 37.4$^\circ$, 
where inelastic scattering events were not observed. (b) Spectrum obtained with a (CH$_2$)$_\text n$ target at a $^7$Be beam energy of 20~MeV and 
$\theta_\text{lab} = $ 29.7$^\circ$, where proton elastic scattering events are well separated from proton inelastic scattering events. $^7$Be+$^{12}$C scattering events are not 
visible here because the gains were set to place the proton scattering data in the middle of the ADC range such that $^7$Be+$^{12}$C scattering events were beyond the range of ADC. }
\label{raw_spectrum} 
\end{center}
\end{figure}

 
For each beam energy, there were two runs, for the purpose of separately collecting $^7$Be+p and $^7$Be+Au events. The two runs were performed with 
a (CH$_2$)$_\text n$ target and a combined target [i.e., a (CH$_2$)$_\text n$ foil with a Au foil in the back], respectively. The proton scattering events could 
be distinguished from the $^7$Be+$^{12}$C scattering events based upon their energies as shown in Fig.~\ref{raw_spectrum} (a).  Proton inelastic scattering events were only observed at high $^7$Be beam energies. The proton inelastic scattering events were well separated from the proton 
elastic scattering events as shown in Fig.~\ref{raw_spectrum} (b).\par

\begin{figure}
\begin{center}
\includegraphics[width=1.0\columnwidth,angle=0]{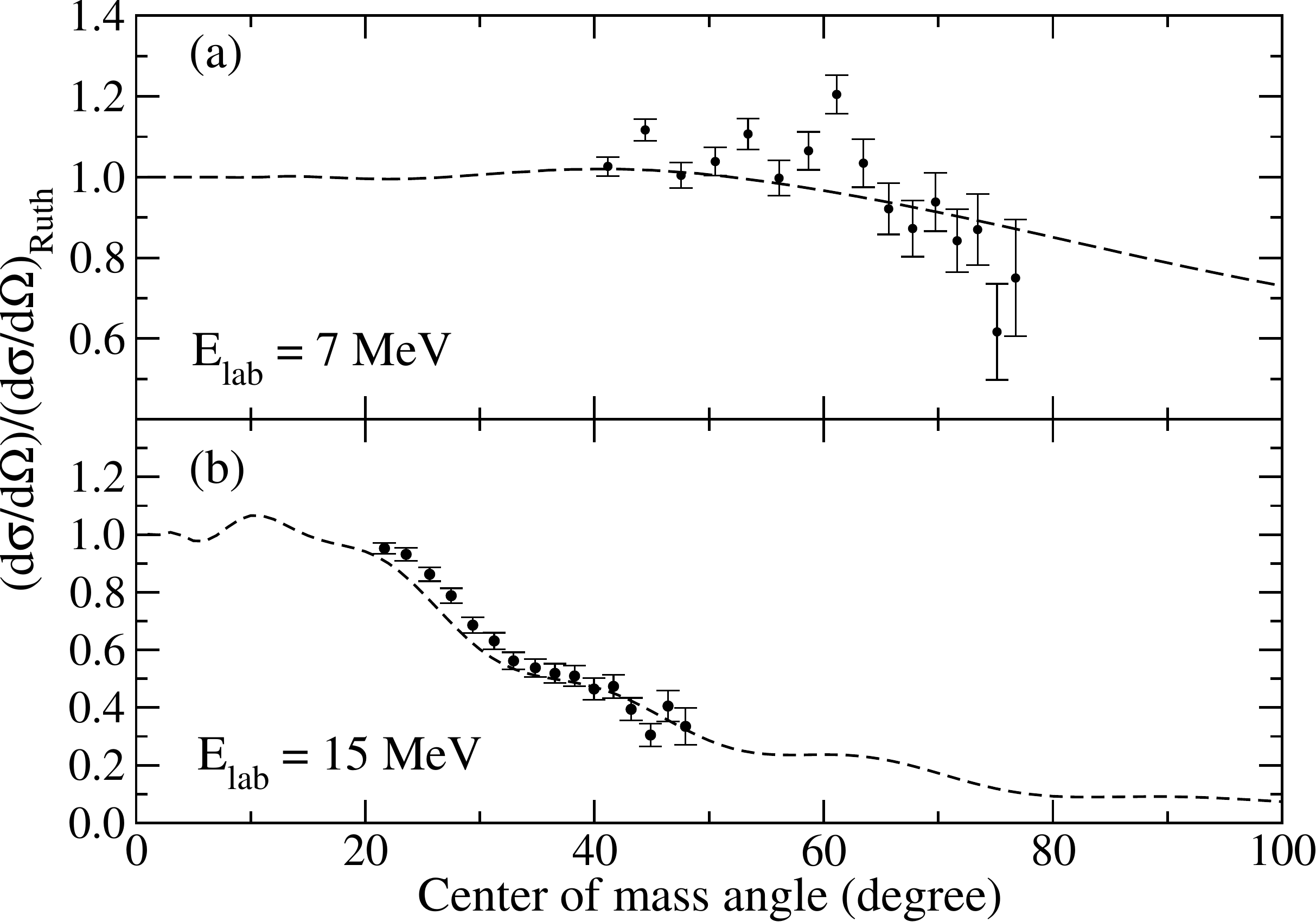}
\caption{$^7$Be+$^{12}$C scattering data from the experiment.  (a) $^7$Be+$^{12}$C scattering data for a $^7$Be beam energy of 7.0~MeV. (b) $^7$Be+$^{12}$C scattering data for a $^7$Be beam energy of 15.0~MeV. Dashed curves are optical model calculations using the parameters 
from Ref.~\cite{Poling_1976}
}
\label{optical_model} 
\end{center}
\end{figure} 
The $^7$Be+p scattering data were normalized to simultaneous scattering reactions. The low energy scattering data (for $^7$Be beam energies of $E_{\text{lab}}$ = 4, 4.5, and 5.2~MeV) 
were normalized to the $^7$Be+$^{12}$C scattering data, as the carbon scattering at these energies is well described by Rutherford scattering. At higher energies,
the $^7$Be+$^{12}$C scattering starts deviating from Rutherford scattering as shown in Fig~\ref{optical_model}. For $^7$Be beam energies of $E_{\text{lab}}$ = 7, 8, 9, 10, 11, 12, 13, and 16~MeV,
the $^7$Be+p scattering data were normalized to $^7$Be+$^{12}$C scattering cross 
sections, which were themselves normalized by $^7$Be+Au scattering data. To utilize this normalization procedure, we need to know the carbon-to-gold ratio rather 
than the absolute target thickness assuming H/C=2. The carbon-to-gold ratio was determined using the ratio of differential cross sections of $^7$Be+$^{12}$C 
and $^7$Be+Au scattering, both of which are described by Rutherford scattering at small angles. The carbon-to-gold ratio was determined to 
be C/Au=10.2$\pm$0.7, where the quoted uncertainty is statistical in nature. For $^7$Be beam energies of $E_{\text{lab}}$ = 19.2 and 22~MeV, the proton scattering data 
were normalized directly to the $^7$Be+Au scattering data, as $^7$Be+Au scattering at all angles and energies covered in this experiment is well described by Rutherford 
scattering. 
For three beam energies ($E_{\text{lab}}$ = 15, 17.5, and 20~MeV), $^7$Be+Au scattering was not measured and $^7$Be+$^{12}$C cross sections were not 
experimentally determined. For these energies the $^7$Be+p scattering was normalized to the $^7$Be+$^{12}$C elastic scattering cross section calculated 
via the optical model using the {\sc DWUCK5} code~\cite{DWUCK5}. The $^7$Li+$^{12}$C optical model parameters from Ref.~\cite{Poling_1972} 
were used to describe $^7$Be+$^{12}$C 
elastic scattering by changing the charge and the incident energy.
This parametrization was found to give a good agreement, to within~10$\%$ of the $^7$Be+$^{12}$C elastic scattering data at energies where the normalization 
was determined independently.\par

The normalization procedures explained before depend on the ratio of the target atoms. The hydrogen-to-carbon ratio in the target was determined from the 4-MeV $^7$Be measurement using the ratio of $^7$Be+p and $^7$Be+$^{12}$C scattering, both of which were assumed to be 
Rutherford scattering. The systematic uncertainty for $^7$Be measurements of $E_{\text{lab}}$ = 4, 4.5, 5.2, 15, 17.5, and 20~MeV, which depends on the hydrogen-to-carbon ratio,
was estimated to be $\pm$6$\%$. For $^7$Be beam energies of $E_{\text{lab}}$ = 7, 8, 9, 10, 11, 12, 13, and 16~MeV, the normalization procedure depends on the carbon-to-gold ratio, 
and the systematic uncertainty was estimated to be $\pm$6$\%$. For measurements at $E_{\text{lab}}$ = 19.2 and 22~MeV, the normalization procedure depends on the hydrogen-to-gold ratio, and the systematic uncertainty
was estimated to be $\pm$7$\%$.
The optical model analysis used for three beam energies ($E_{\text{lab}}$ = 15, 17.5, and 20~MeV) has an additional systematic uncertainty of $\pm$7$\%$, thus the 
overall systematic uncertainty for these energies was estimated to be $\pm$10$\%$. \par

\begin{figure}
\begin{center}
\includegraphics[width=1.0\columnwidth,angle=0]{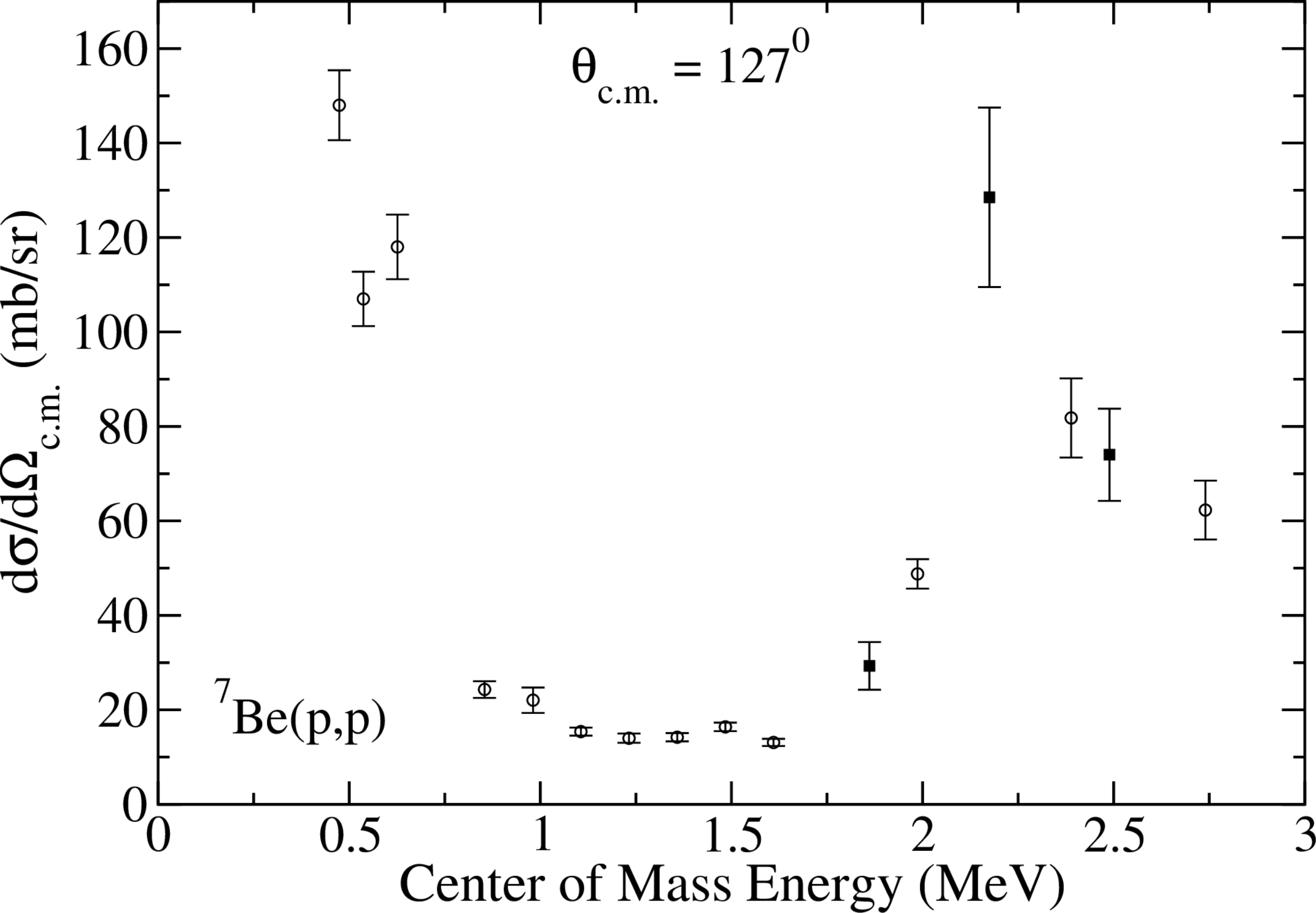}
\caption{Excitation function for $^7$Be+p elastic scattering at $\mathrm{\theta}_\text{c.m.}$=127$^\circ$. Circles and squares correspond
 to the data from the first and second experimental campaigns, respectively.}
\label{raw_excitation} 
\end{center}
\end{figure}

Figure~\ref{raw_excitation} shows the excitation function for elastic scattering of $^7$Be+p measured in this work. Circles and squares
correspond to the data from the first and second experimental campaigns, respectively.

\section{\textit{R}-MATRIX ANALYSIS}
\label{R-matrix} 
The differential scattering cross section for $^7$Be(p,p)$^7$Be is described using \textit{R}-matrix theory~\cite{Lane_Thomas}. The elastic 
and inelastic cross-section data from this experiment and low-energy elastic scattering data from Angulo \textit{et al}.~\cite{ANGULO2003211} 
have been analyzed using the multilevel multichannel code AZURE2~\cite{AZURE2}. The alternative parametrization of the \textit{R}-matrix theory presented 
in Ref.~\cite{Brune_2002} is used. So, the $R$-matrix can be expressed in terms of alternative parameters namely the observed resonance 
energy $\tilde{E}$ and the observed reduced width amplitude ${\tilde{\gamma}}$. A channel radius of 4.3~fm is assumed and the background poles have been fixed at particular 
excitation energies. \par
\begin{figure}
\begin{center}
\includegraphics[width=1.0\columnwidth,angle=0]{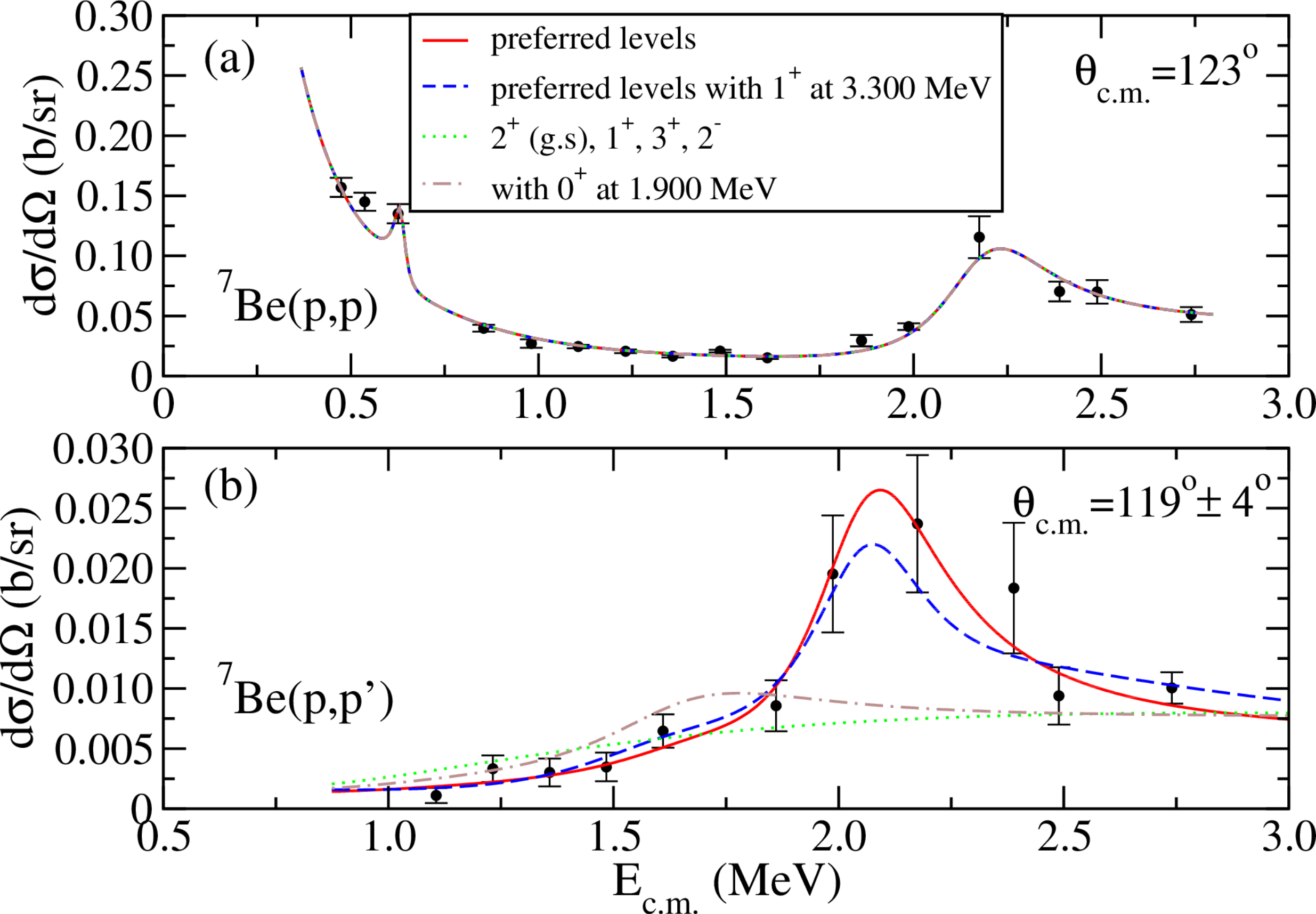}
\caption{ \textit{R}-matrix fit of elastic and inelastic scattering data from this work. The dotted green curve corresponds to the fit obtained with 2$^+$, 
1$^+$, 3$^+$, and 2$^-$ levels at 0, 0.77, 2.32, and 3.52~MeV, respectively. The dashed-dotted brown curve corresponds to the fit obtained with an additional 0$^+$ level at 1.9~MeV. 
The dashed blue curve corresponds to the fit with the preferred levels with an additional 1$^+$ level at 3.3~MeV
 and the solid red curve corresponds to the fit with the preferred levels only.
\label{states}}
\end{center}
\end{figure}

The spins of the ground and first excited states of $^7$Be are 3/2$^-$ and 1/2$^-$, respectively. If we restrict our calculations up to p waves, then the allowed 
levels in $^8$B following the coupling scheme would be 0$^-$, 0$^+$, 1$^-$, 1$^+$, 2$^-$, 2$^+$, and 3$^+$. 
The \textit{R}-matrix analysis was started with the states of $^8$B identified in previous experiments~\cite{ANGULO2003211, Goldberg1998, Rogachev_2001, 
Yamaguchi2009, Mitchell_2013}, namely the 2$^+$, 1$^+$, 3$^+$ and 2$^-$ levels at excitation energies of 0, 0.77, 
2.32, and 3.52~MeV, respectively. The separation energies for the levels introduced in the $R$-matrix analysis were taken from Ref.~\cite{ENSDF}. The values of the asymptotic normalization constants (ANC) used for the ground state in this analysis are $C{^2}_{(^{3}P_{2})}$=0.0990(57)~fm$^{-1}$,
$C{^2}_{(^{5}P_{2})}$=0.438(23)~fm$^{-1}$, and $C{^2}_{(^{3}P^{*}_{2})}$=0.1215(36)~fm$^{-1}$~\cite{Zhang:2017}, where the third value refers to the $^7$Be excited state component and the ANC's were obtained using \textit{ab initio} methods ~\cite{Nollet_Wiringa}. The fits to the scattering data is not highly sensitive to the choice of ANC values in this analysis.
These states reproduce the fits to the elastic scattering data reasonably well, as shown in Fig.~\ref{states}, but could not explain the inelastic scattering data.
In Fig.~\ref{states}(b), data points correspond to the inelastic scattering cross section for a center-of-mass angle 119$^{\circ}\pm$4$^{\circ}$.
The conversion from laboratory angle to center-of-mass angle was done taking into account the correct kinematics for inelastic scattering.
Under the assumption of just the known literature values the inelastic channel was not well reproduced, so alternative level schemes were used for the $R$-matrix parameters
in order to improve the fit. 
Additional 0$^+$, 1$^-$, and 2$^+$ states at excitation energies of 1.9, 9.0, and 2.21~MeV were introduced to  improve the fits to the inelastic
scattering data with no significant changes in the fits to the elastic scattering data. 
The 0$^+$ level at an excitation energy of 1.9 MeV in $^8$B was previously suggested in Ref.~\cite{Mitchell_2013}. The 1$^-$ level is introduced as a background level in our fits. 
In the phenomenological \textit{R}-matrix theory, levels introduced at energies higher than the highest energy data points and 
with large widths are termed background levels. The solid red line in Fig.~\ref{states} represents the fit with all these levels. 
These levels are defined as preferred levels hereafter. It can be infered  from Fig.~\ref{states} that the 2$^+$
level at 2.21~MeV is required to fit the inelastic scattering data well. The introduction of an additional 2$^-$ level at 9.0 MeV as a background level does not change significantly the fits to the data, so it was not included in our final fit. The sensitivity of the fit to the excitation energy of the 2$^+$ level was studied and we differ in the extracted excitation energy for such a level from Ref.~\cite{Mitchell_2013}. 

\begin{figure}
\begin{center}
\includegraphics[width=1.0\columnwidth,height=6.0cm,angle=0,keepaspectratio]{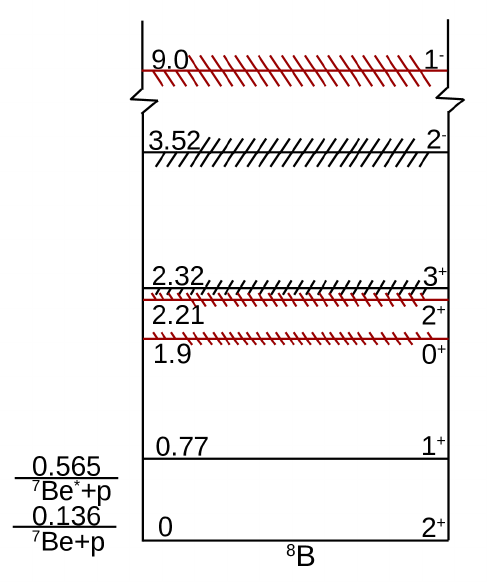}
\caption{ Level structure for $^8$B. This work supports the existence of states shown in red that have been previously suggested in 
Ref.~\cite{Mitchell_2013}.
\label{level_structure}}
\end{center}
\end{figure}

The existence of a 1$^+$ level around 2 to 3~MeV in $^8$B has often been questioned. Gol'dberg \textit{et al}.~\cite{Goldberg1998} suggested a 1$^+$ level at
2.83$\pm$0.150~MeV with a width of 780$\pm$200~keV. Mitchell \textit{et al}.~\cite{Mitchell_2013} introduced a 1$^+$ level at 3.3~MeV with
a width of 2.8~MeV. The recoil corrected continuum shell-model calculations in Ref.~\cite{Halderson} also suggested the presence of a 1$^+$ level 
in $^8$B requiring verification by inelastic scattering measurements. The dashed blue curve in Fig.~\ref{states} shows the effect of a 1$^+$ 
level at an excitation energy of 3.3~MeV along with the preferred levels. The fits to the data with and without this 1$^+$ level can be compared in Fig.~\ref{states}. 
There is no significant
change in the elastic excitation function but the inelastic scattering cross section is underestimated. Therefore, based on the scattering data available for $^7$Be+p, there is no conclusive evidence for a 1$^+$ level at an excitation energy of 3.3~MeV. 
 Based on the analysis of these data, the level structure of the $^8$B is shown in Fig.~\ref{level_structure}. \\

 \par

 \subsection{Scattering length from \textit{R}-matrix analysis}
 In this section, we relate the $s$-wave scattering lengths to the best-fit \textit{R}-matrix parameters. The collision matrix $U_{c'c}$ can be expressed 
 as
 
 \begin{equation}\label{collision}
  U_{c'c}=\Omega_{c'}\Omega_{c}[\delta_{c'c}+2i(P_{c'}P_{c})^{1/2}M_{c'c}],
 \end{equation}
where $M_{c'c}=\mathbf{\tilde{\gamma}}^{T}_{c'}\mathbf{\tilde{A}}\mathbf{\tilde{\gamma}}_{c}$. ${\mathbf{\tilde{A}}}$ is the level matrix as defined in Ref.~\cite{Brune_2002}, $P_{c}$ is the penetration factor, and $c$ is the channel index. For single-channel elastic scattering Eq.~(\ref{collision})
reduces to 
\begin{equation}
  U_{cc}=\Omega_{c}^{2}[1+2iP_{c}(E)M_{cc}],
  \end{equation}
where 
 \begin{equation}
  \Omega_{c}=e^{i(\omega_{c}-\phi_{c})}.
  \end{equation} 
The quantities $\phi_{c}$ and $\omega_{c}$ are the hard sphere phase shift and Coulomb phase shift respectively. 
For $s$-wave scattering ($l$=0), $\omega_{c}=0$. For diagonal collision matrix elements, $U_{cc}=e^{2i \delta_{c}}$, where $\delta_{c}$ 
is the total phase shift. In this case, the phase shift can be related to $R$-matrix parameters via 
\begin{equation}\label{phase_shift}
   e^{2i\delta_{c}}=e^{-2i\phi_{c}}\left[1+2iP_{c}(E)M_{cc}\right].
  \end{equation}
In the low-energy limit, Eq.~(\ref{phase_shift}) can be written as

\begin{equation}\label{cotphaseshift}
  \lim_{k\to 0} \cot\delta_{0}=\frac{1}{-\phi_{0}+P_{0}M_{cc}},
   \end{equation}
  with 
   $P_{0}=ka/G_{0}^{2}(ka)$; $a$ is the channel radius, $k$ is the wave number, and $G_{0}$ is the irregular Coulomb function for $l$=0. 
In the limit $k\to 0$, the effective range expansion from~\cite{Bethe1949} can be reduced to
\begin{equation}\label{effective_range}
 \lim_{k\to 0} \left[k C_{0}^2 \cot \delta_{0}\right]=-\frac{1}{a_{0}},
 \end{equation} 
 where $C_{0}^2=2\pi\eta/(e^{2\pi\eta}-1)$, with $\eta$ the Sommerfeld parameter. 
From Eq.~(\ref{cotphaseshift}) and Eq.~(\ref{effective_range}), the expression for the $s$-wave scattering length ($a_{0})$ in terms of $R$-matrix parameters is obtained as

\begin{equation}\label{scatt_length}
  a_{0}=-a\left[\frac{M_{cc}}{x^{2}K_{1}^{2}(x)}-\frac{2I_{1}(x)}{x^{2}K_{1}(x)}\right],
 \end{equation}

  where $I_{1}(x)$ and $ K_{1}(x)$ are modified Bessel functions and $x=\left(8 Z_{1}Z_{2}e^{2}\mu a /\hbar^{2} \right)^{1/2}$, 
  $Z_{1}e$ and $Z_{2}e$ are the nuclear charges, $\hbar$ is the reduced Planck's constant, $\mu$ is the reduced mass, and $a$ is the $R$-matrix channel radius. The Coulomb functions
 have been expressed in terms of modified Bessel functions using Ref.~\cite{Humblet_1985}.

\section{RESULTS}
\label{results}
The elastic and inelastic angular distribution data from the ORNL measurement and the elastic scattering data from Ref.~\cite{ANGULO2003211} have been fitted simultaneously. The 
low-energy data from Ref.~\cite{ANGULO2003211} were introduced to constrain the fits below 1~MeV center-of-mass energy. The systematic uncertainties of 
both data sets were introduced in the simultaneous fitting. In AZURE2, the systematic
uncertainty for the data is introduced in the normalization of the data. A systematic
uncertainty of $\pm$5.5$\%$ has been assumed for the data from Ref.~\cite{ANGULO2003211} as quoted in the paper, while the systematic uncertainties for different angular distributions 
from the ORNL measurement are included as explained in Sec.~\ref{experiment}. The absolute normalization of the data is allowed to vary during the fits. The 
output of the fit along with the chi-square values for each data segment are presented in Table~\ref{chi-square}. 
The best-fit parameters from the simultaneous fitting are presented in Table~\ref{R-matrix_param}. \par

\begin{figure}
\begin{center}
\includegraphics[width=1.0\columnwidth,angle=0]{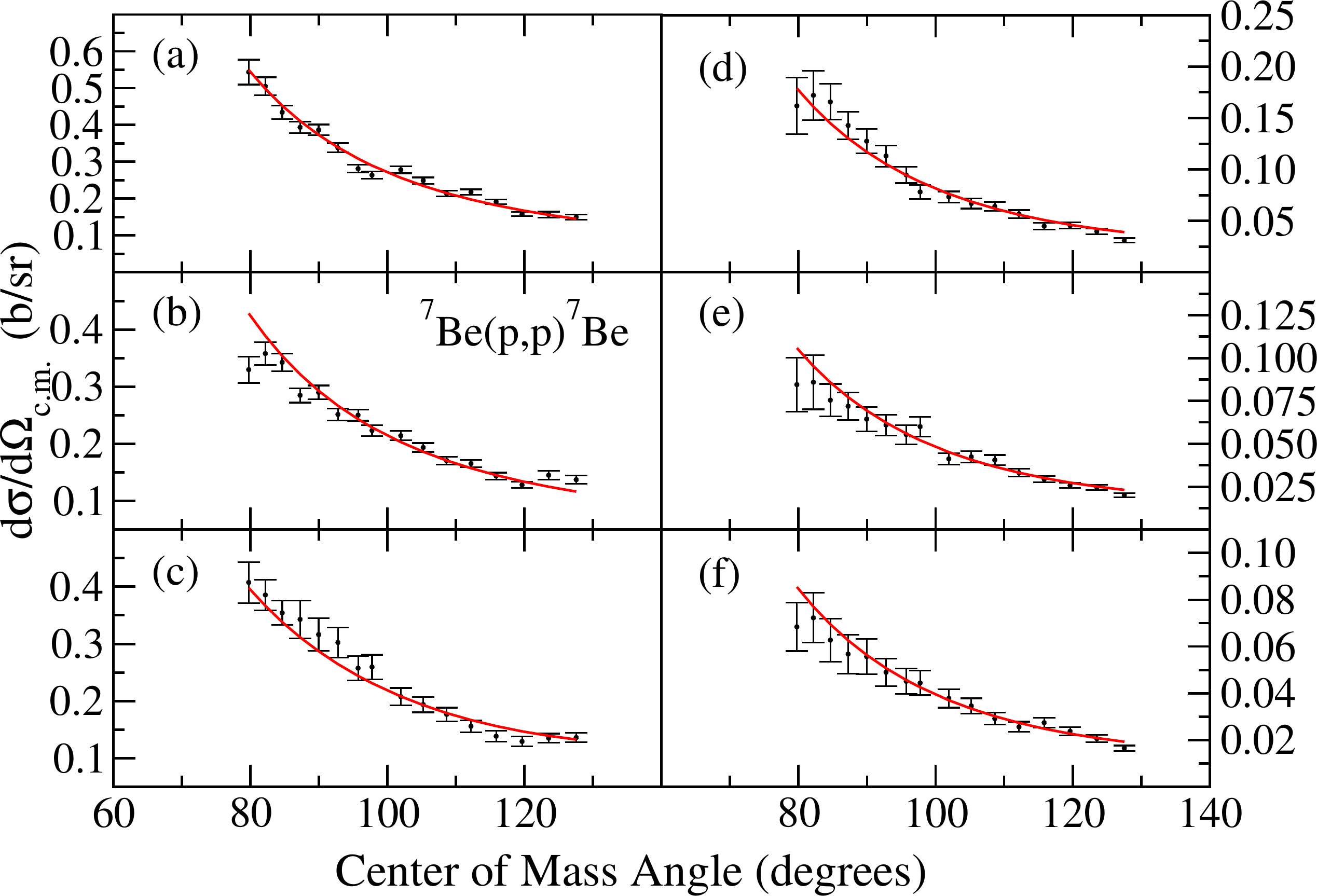}
\caption{Fits to the $^7$Be(p,p)$^7$Be angular distribution data from this work at (a) E$_{\text{c.m.}}$=0.474~MeV, (b) E$_{\text{c.m.}}$=0.537~MeV, (c) E$_{\text{c.m.}}$=0.626~MeV, (d) E$_{\text{c.m.}}$=0.854~MeV, (e) E$_{\text{c.m.}}$=1.106~MeV, and (f) E$_{\text{c.m.}}$=1.232~MeV.
\label{ornl_elastic_1}}
\end{center}
\end{figure}

\begin{figure}
\begin{center}
\includegraphics[width=1.0\columnwidth,angle=0]{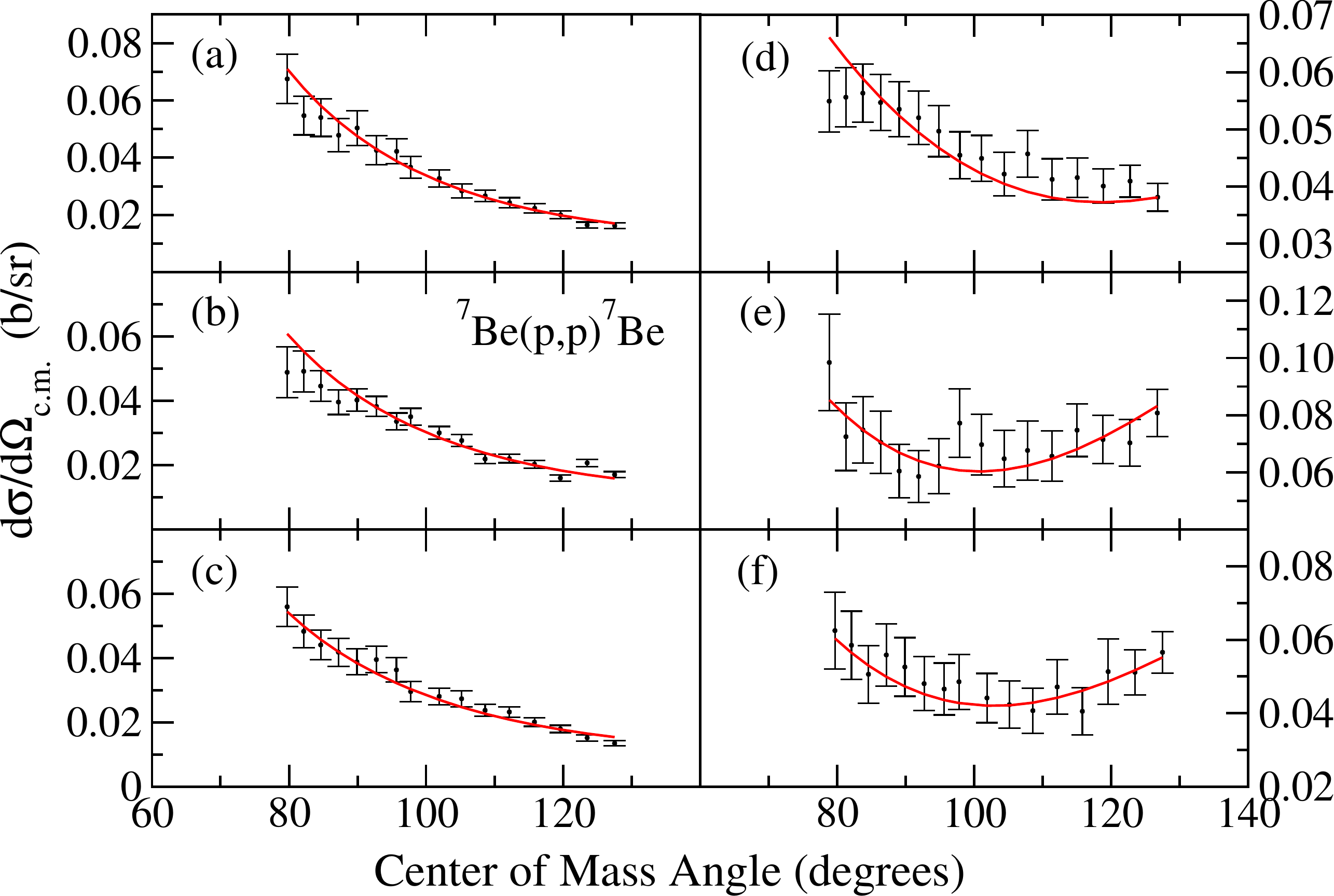}
\caption{Fits to the $^7$Be(p,p)$^7$Be angular distribution data from this work at (a) E$_{\text{c.m.}}$= 1.358~MeV, (b) E$_{\text{c.m.}}$=1.484~MeV, (c) E$_{\text{c.m.}}$=1.610~MeV, (d) E$_{\text{c.m.}}$=1.987~MeV, (e) E$_{\text{c.m.}}$=2.389~MeV, and (f) E$_{\text{c.m.}}$=2.740~MeV.
\label{ornl_elastic_2}}
\end{center}
\end{figure}

\begin{figure}
\begin{center}
\includegraphics[width=1.0\columnwidth,angle=0]{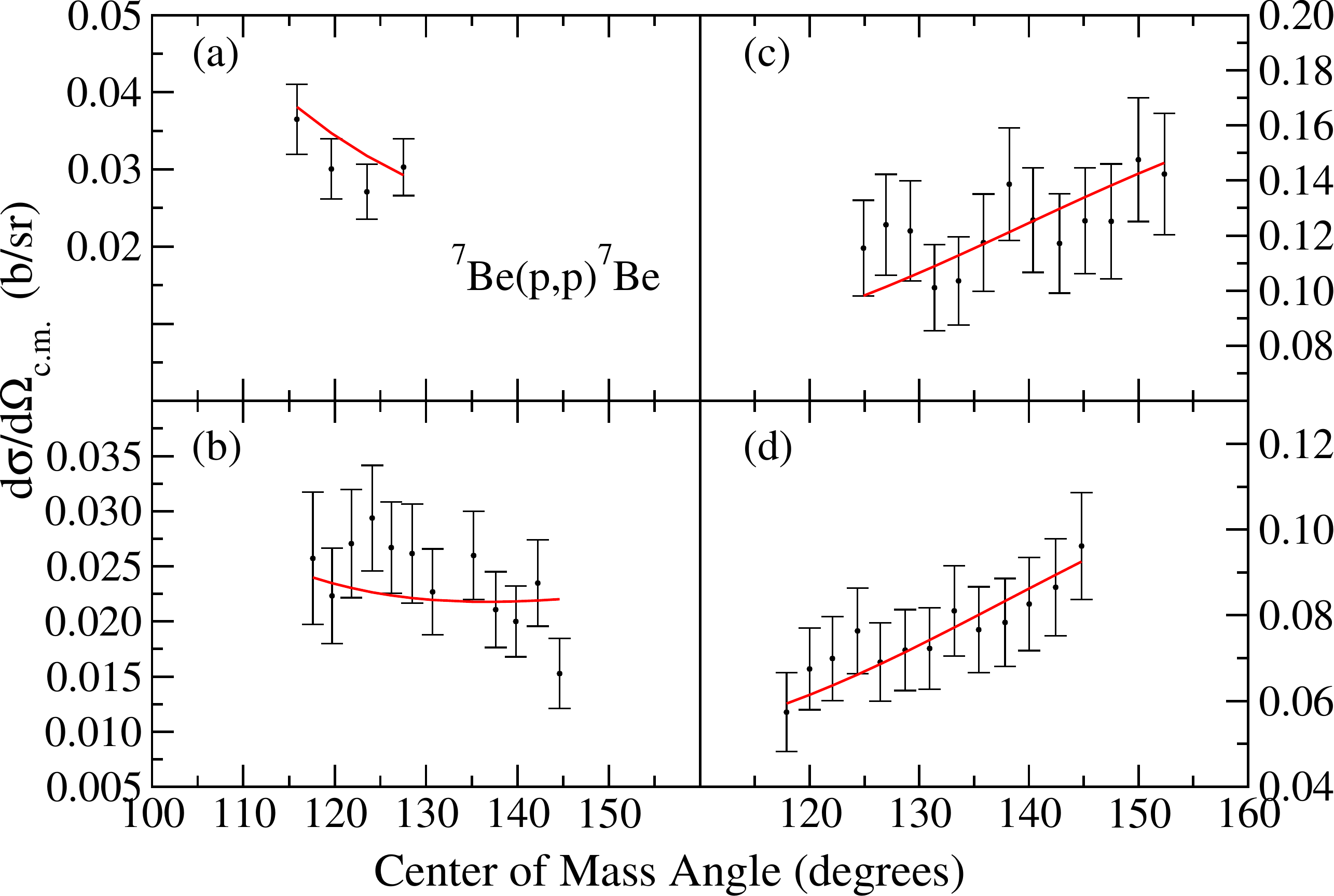}
\caption{Fits to the $^7$Be(p,p)$^7$Be angular distribution data from this work at (a) E$_{\text{c.m.}}$=0.981~MeV,  (b) E$_{\text{c.m.}}$=1.861~MeV, (c) E$_{\text{c.m.}}$= 2.175~MeV, and (d) E$_{\text{c.m.}}$= 2.489~MeV.
\label{ornl_elastic_3}}
\end{center}
\end{figure}

\begin{figure}
\begin{center}
\includegraphics[width=1.0\columnwidth,angle=0]{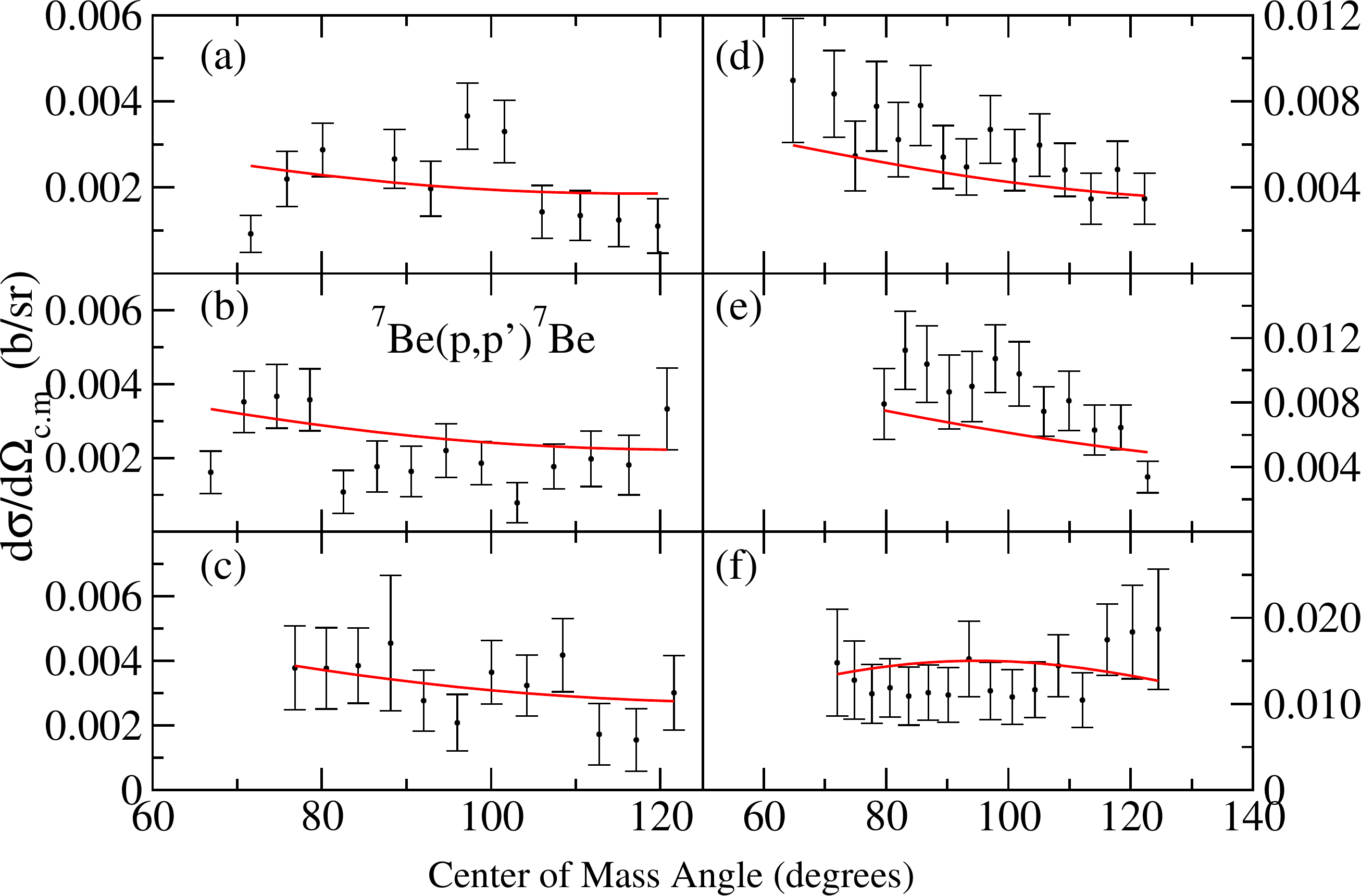}
\caption{Fits to the $^7$Be(p,p')$^7$Be angular distribution data from this work at (a) E$_{\text{c.m.}}$=1.106~MeV, (b) E$_{\text{c.m.}}$=1.232~MeV, (c0 E$_{\text{c.m.}}$=1.358~MeV, (d) E$_{\text{c.m.}}$=1.484~MeV, (e) E$_{\text{c.m.}}$=1.610~MeV, and (f) E$_{\text{c.m.}}$=2.389~MeV.
\label{ornl_inelastic_1}} 
\end{center}
\end{figure}

\begin{figure}
\begin{center}
\includegraphics[width=1.0\columnwidth,angle=0]{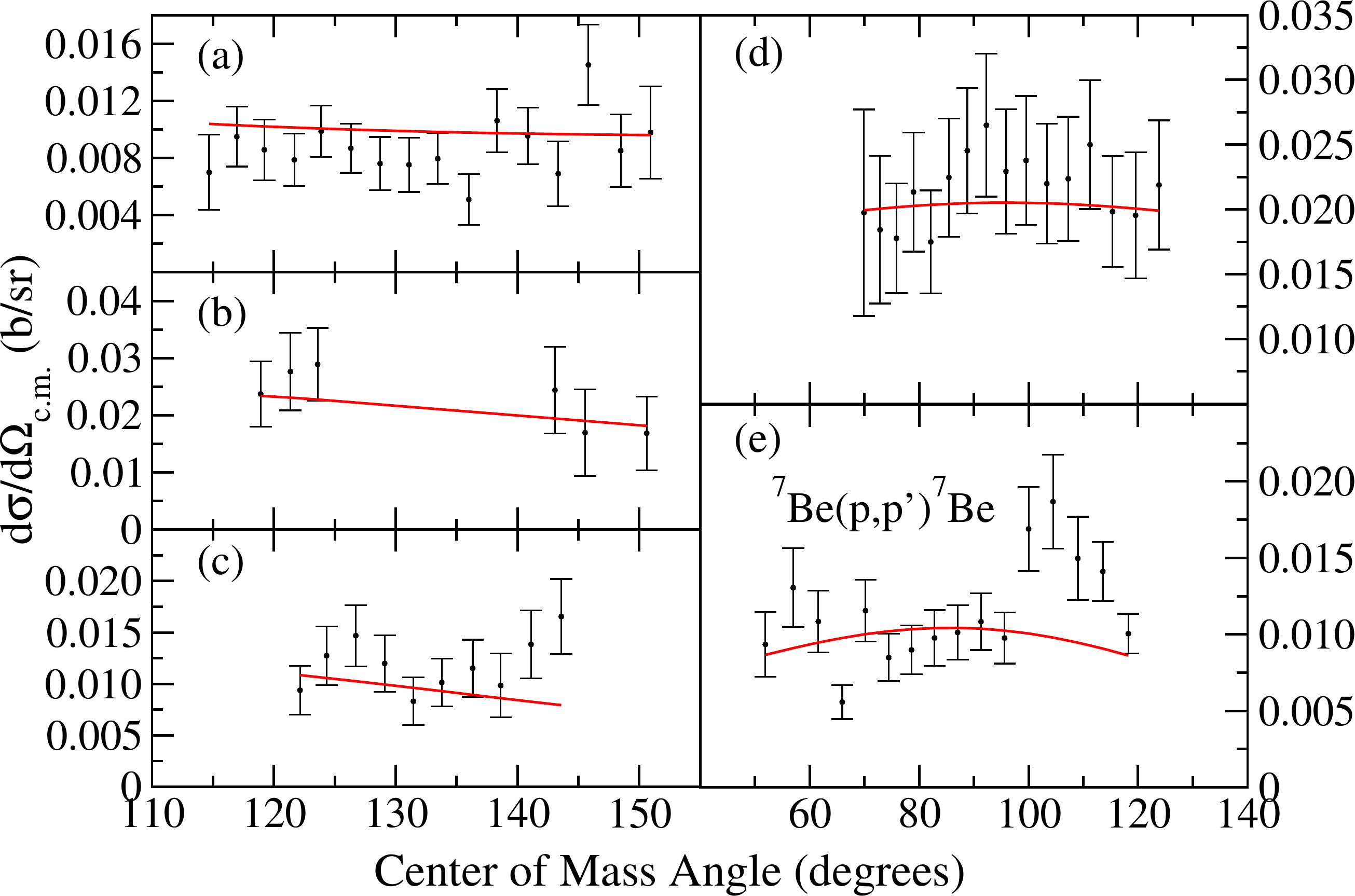}
\caption{Fits to the $^7$Be(p,p')$^7$Be angular distribution data from this work at (a) E$_{\text{c.m.}}$=1.861~MeV, (b) E$_{\text{c.m.}}$=2.175~MeV, (c) E$_{\text{c.m.}}$=2.489~MeV, (d) E$_{\text{c.m.}}$=1.987~MeV, and (e) E$_{\text{c.m.}}$=2.74~MeV.
\label{ornl_inelastic_2}}
\end{center}
\end{figure}

\begin{figure}[ht] 
\begin{center}
\includegraphics[width=1.0\columnwidth,angle=0]{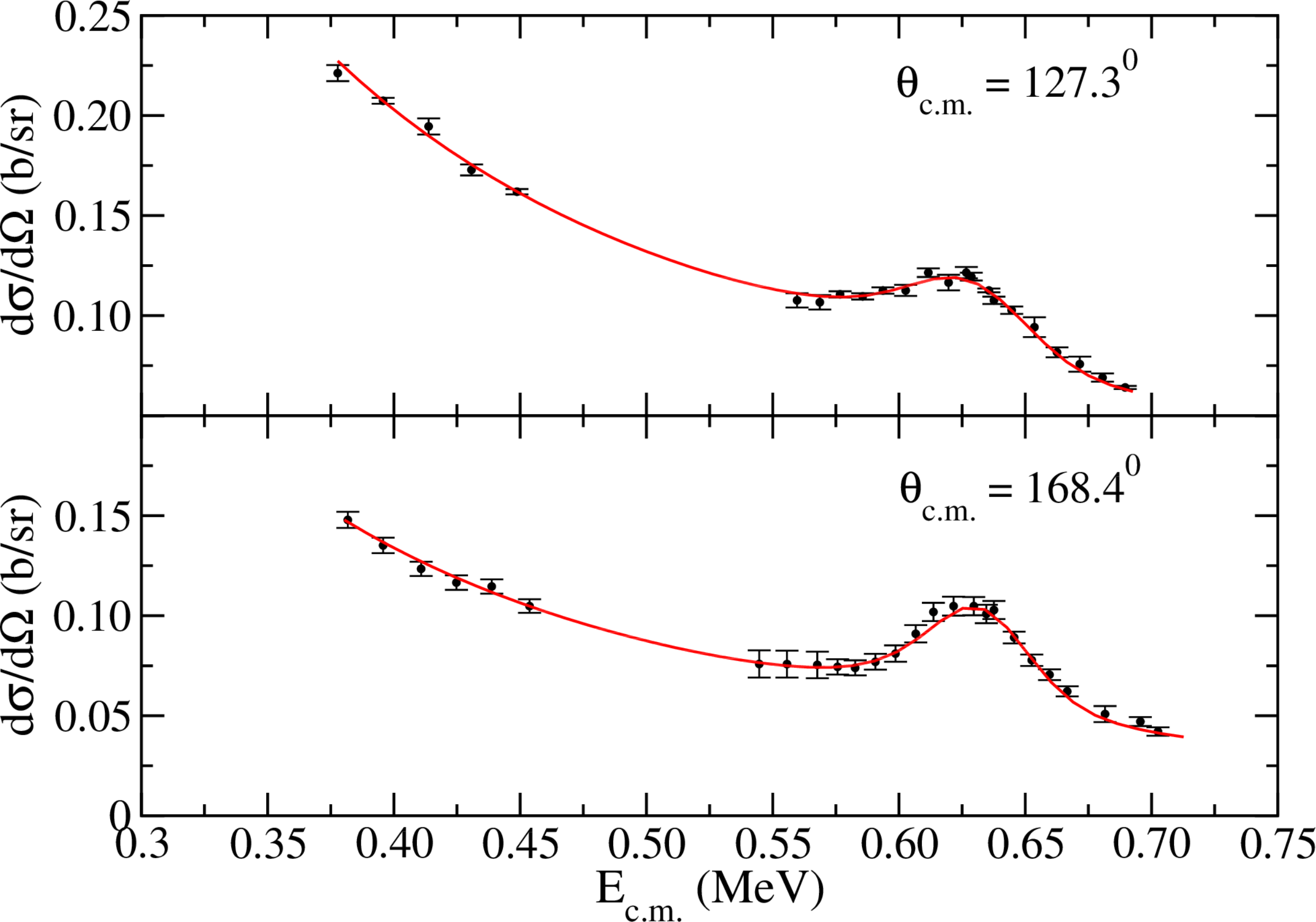}
\caption{ $^7$Be+p elastic scattering excitation function at low energies from Ref.~\cite{ANGULO2003211}. 
 The best fit is shown by the solid red curve, and the filled black circles represent the data. The experimental cross section was convoluted to account for the 14- and 19-keV experimental resolutions reported in~\cite{ANGULO2003211}.
 A systematic uncertainty of $\pm$5.5$\%$ was included in the calculation.
\label{angulo}}
\end{center}
\end{figure}

The fits to the elastic angular distributions are presented in Figs.~\ref{ornl_elastic_1}, ~\ref{ornl_elastic_2}, and~\ref{ornl_elastic_3} and the 
fits to the inelastic angular distributions are presented in Fig.~\ref{ornl_inelastic_1} and Fig.~\ref{ornl_inelastic_2}. 
The fits to data from Ref.~\cite{ANGULO2003211} are shown in Fig.~\ref{angulo}. \par 

Using the best-fit parameters from Table~\ref{R-matrix_param} and Eq.~(\ref{scatt_length}), the $s$-wave scattering lengths for channel spins 1 and 2 were 
calculated to be 17.34$^{+1.11} _{-1.33}$ and -3.18$^{+0.55}_{-0.50}$~fm, respectively. 
To illustrate the sensitivity of the fit to the reduced width amplitudes of the 1$^-$ and 2$^-$ levels, the reduced width amplitudes for these levels were varied and the 
changes in the total $\chi^2$ were compared. A change of $\Delta \chi^2$=1 is used to define the acceptable range of the reduced width amplitudes for these levels, 
which gives the error bars in the scattering length values for channel spins 1 and 2, respectively. Using the same approach, the 1-$\sigma$ error bar was estimated for
the parameters of the 0$^+$ and 2$^+$ levels. The widths of the 0$^+$ level are $\Gamma_{\text{p}}$=0.120$\pm$0.028 MeV and $\Gamma_{\text{p'}}$=0.428$\pm$0.130 MeV, where
$\Gamma_{\text{p}}$ and $\Gamma_{\text{p'}}$ refer to the elastic and inelastic channel widths, respectively. Similarly, the widths of the 2$^+$ level are $\Gamma_{\text{p}}$=0.024$\pm$0.009 MeV and 
$\Gamma_{\text{p'}}$=0.230$\pm$0.001 MeV. The excitation energies of the 0$^+$ and 2$^+$ levels are 1.9$\pm$0.1 and 2.21$\pm$0.04 MeV, respectively. Our excitation energy for the 2$^+$ level differs from the value presented in Ref.~\cite{Mitchell_2013}. The elastic proton partial width for the 1$^+$ state (0.77 MeV) from our analysis is in agreement with the value reported in Ref.~\cite{ANGULO2003211}.

Table~\ref{chi-square} lists the $\chi^2$ values of the fit to each data set.
 The fits to the first two data segments in both the elastic and the inelastic scattering from this work have a large $\chi^2$. There are no obvious reasons for this, but
point-to-point uncertainty is one possible explanation. Sensitivity tests were performed by excluding segments
with large $\chi^2$ values (i.e., $\chi^2$/N \textgreater 2) and segments with normalization factors above or below 20$\%$ (i.e, Norm \textless 0.80 and Norm \textgreater 1.20). 
Excluding segments with $\chi^2$/N \textgreater 2 does not affect the normalizations of the included segments considerably. Similar conclusions were obtained by excluding the segments
following the normalization criterion. 
Also, the data from Ref.~\cite{ANGULO2003211} were fitted alone, starting with the parameters in Table~\ref{R-matrix_param}, to evaluate the effects on the scattering
length values. If the well-known states [2$^+$ (ground state), 1$^+$ (0.77~MeV), and 3$^+$ (3.52~MeV)] alone are included to fit the data from Ref.~\cite{ANGULO2003211} along with
the 2$^-$ and 1$^-$ background levels, we obtain $s$-wave scattering lengths consistent with the results in Ref.~\cite{ANGULO2003211}. But with the introdution of the
inelastic channel along with the inclusion of the 0$^+$ (1.9-MeV), and 2$^+$ (2.21-MeV) states, the results for the $s$-wave scattering lengths differ significantly from the results 
in Ref.~\cite{ANGULO2003211}.
The scattering lengths obtained from this analysis along with the values published in the literature are presented in Table~\ref{Scattering_length}. Angulo~\textit{et al}. 
made the only previous determination of $s$-wave scattering lengths for the $^7$Be+p system, where the cross-section data have been analyzed in an $R$-matrix framework and the $s$-wave
scattering lengths have been deduced. Navratil~\textit{et al}.~\cite{NAVRATIL2011} used the \textit{ab initio} no-core shell model/resonating group
method to calculate the $^7$Be(p,$\gamma$)$^8$B radiative capture cross section and deduce the $s$-wave scattering lengths for $^7$Be+p. The $s$-wave scattering lengths from Ref.~\cite{NAVRATIL2011} 
do not agree with the results of this analysis.\par

\par
In the context of the potential model~\cite{Baye_2000}, the extrapolation of $S_{17}$ down to solar energies depends on the value of the average scattering length ($\bar{a_{0}}$), defined as 
\begin{equation}\label{avg_scatt_length}
\bar{a_{0}}= \frac{C^{2}_{{(^{3}P_{2}})} a_{01} + C^{2}_{{(^{5}P_{2})}} a_{02}}{C^{2}_{(^{3}P_{2})}+C^{2}_{(^{5}P_{2})}}. 
\end{equation}
The $\bar{a_0}$ value deduced from this work is
0.60$^{+0.15}_{-0.18}$~fm using the ANC values from Ref.~\cite{Zhang:2017} neglecting their uncertainties. This shows that the average scattering length can be better constrained than the individual scattering lengths for channel spins 1 and 2, respectively.

\begin{table}
\caption{Normalization factors, $\chi^2$, and number of data points ($N$) for angular distributions of the ORNL measurement (Parts A and B)  and excitation function from Ref.~\cite{ANGULO2003211} (part C). The energies and the angles are in the center of mass frame.}
\begin{ruledtabular}
\begin{tabular}{llll}
Reaction ($E_{\text{c.m.}}/\theta_{\text{c.m.}}$)  & Norm & ~~$\chi^{2}$ & N\\
\hline
  & (A) $^{7}$Be(p,p)$^{7}$Be &  &\\
0.474~MeV & 1.012 & 37.750 & 16\\
0.537~MeV & 1.283 & 53.301 & 16\\
0.626~MeV & 1.039 & 17.119 & 16\\
0.854~MeV & 1.274 & 24.176 & 16\\
0.981~MeV & 1.375 & 3.333  & 4\\
1.106~MeV & 1.300 & 18.982 & 16\\
1.232~MeV & 1.177 & 12.599 & 16\\
1.358~MeV & 1.144 & 8.223 & 16\\
1.484~MeV & 1.039 & 29.985 & 16\\
1.610~MeV & 1.032 & 14.725 & 16\\
1.861~MeV & 0.893 & 10.891 & 12\\
1.987~MeV & 0.781 & 15.205 & 16\\
2.175~MeV & 0.964 & 5.815 & 13\\
2.389~MeV & 0.987 & 6.270 & 16\\
2.489~MeV & 0.933 & 2.706 & 13\\
2.740~MeV & 0.908 & 3.351 & 16\\

 &  (B) $^{7}$Be(p,p')$^{7}$Be(1/2$^-$) &  &\\

1.106~MeV & 1.060 & 27.143 & 11\\
1.232~MeV & 1.177 & 34.612 & 14\\
1.358~MeV & 1.022 & 6.982 & 12\\
1.484~MeV & 0.871 & 13.540 & 15\\
1.610~MeV & 0.858 & 21.685 & 12\\
1.861~MeV & 1.398 & 20.647 & 16\\
1.987~MeV & 0.939 & 5.095  & 16\\
2.175~MeV & 0.980 & 2.020 & 6\\
2.389~MeV & 1.120 & 12.803 & 16\\
2.489~MeV & 0.930 & 13.521 & 10\\
2.740~MeV & 0.713 & 46.574 & 16\\

 &  (C) $^{7}$Be(p,p)$^{7}$Be  &  &\\

120.24$^\circ$-131.13$^\circ$ & 0.987 & 98.966 & 87\\
156.62$^\circ$-170.21$^\circ$ & 0.978 & 236.707 & 343\\
\end{tabular}
\end{ruledtabular}
\label{chi-square}
\end{table}

\begin{table}
\caption{ Observed energies and reduced width amplitudes obtained from the best \textit{R}-matrix fit with the channel radius set at 4.3~fm. States with excitation energy in the parentheses are introduced as background 
levels. Parameters values in boldface were treated as fit parameters and all others were held constant. The observed partial widths can be computed from the reduced width amplitudes using Eq.~(41) of Ref.~\cite{AZURE2}.}
\begin{ruledtabular}
\begin{tabular}{llllll}
\textit{J$^{\pi}$} & $\tilde{E}_{x}$ & \textit{$\tilde{\gamma}_{el}$} S=1 & \textit{$\tilde{\gamma}_{el}$} S=2 & \textit{$\tilde{\gamma}_{inl}$} S=0 &\textit{$\tilde{\gamma}_{inl}$} S=1 \\
& (MeV)& (MeV$^{\frac{1}{2}}$) & (MeV$^{\frac{1}{2}}$)& (MeV$^{\frac{1}{2}}$) & (MeV$^{\frac{1}{2}}$)\\
\hline
2$^{+}$ & 0.000 & -0.456 &-0.959 & 0.000 & 0.510\\
1$^{+}$& 0.774 &  \textbf{1.484} & \textbf{-0.268} &\textbf{-0.004} & \textbf{2.904}\\
0$^{+}$ & 1.900 & \textbf{0.501} & 0.000 & 0.000 & \textbf{1.201}\\
2$^{+}$ & 2.210 & \textbf{-0.274} & \textbf{0.323} & 0.000 & \textbf{0.632}\\
3$^{+}$ & 2.320 & 0.000 & 0.607 & 0.000 & 0.000\\
2$^{-}$& 3.520 &  0.000 & \textbf{1.700} & 0.000 & 0.000\\
1$^{-}$& (9.000)  & \textbf{1.433} & 0.000 & 0.000 & \textbf{-1.822}\\
2$^{+}$& (9.000) & \textbf{-77.322} &\textbf{-332.657} & 0.000 &\textbf{66.565}\\
3$^{+}$ & (14.000)& 0.000 & \textbf{1.514} & 0.000 & 0.000\\
\end{tabular}
\end{ruledtabular}
\label{R-matrix_param}
\end{table}

\begin{table}
\caption{$s$-wave scattering lengths for the $^7$Be+p system.}
\begin{ruledtabular}
\begin{tabular}{lll}
$a_{01}$ (fm)& $a_{02}$ (fm) & Reference\\
\hline
17.34$^{+1.11}_{-1.33}$ & -3.18$^{+0.55}_{-0.50}$ & This work\\
25$\pm$9 & -7$\pm$3 & Angulo~\textit{et al}.~\cite{ANGULO2003211}\\
 -5.2& -15.3 & Navratil~\textit{et al}.~\cite{NAVRATIL2011}\\
\end{tabular}
\end{ruledtabular}
\label{Scattering_length}
\end{table}

\section{CONCLUSIONS}
\label{conclusion}
The angular distributions for $^7$Be+p elastic and inelastic scattering were measured in the center-of-mass energy range 0.474-2.740~MeV and center-of-mass
angular range 70$^\circ$-150$^\circ$. Simultaneous fits of the angular distributions from this measurement and the excitation functions from Ref.~\cite{ANGULO2003211}
indicate the existence of a 0$^+$ state at 1.9~MeV and a 2$^+$ state at 2.21~MeV in $^8$B. These states are required to explain the inelastic scattering excitation
function, which shows a clear peak at 2.2~MeV. The results of this analysis do not provide conclusive evidence for the existence of a 1$^+$ level at 3.3~MeV in $^8$B. \par

The experimental determination of $s$-wave scattering lengths for the $^7$Be+p system from an \textit{R}-matrix analysis of elastic and inelastic 
scattering data has been presented. The scattering
length for channel spin 1 is in agreement with the previously reported scattering length in Ref.~\cite{ANGULO2003211}. 
Our result for channel spin 2 lies just outside the 1-$\sigma$ lower limit of the scattering length reported in Ref.~\cite{ANGULO2003211}. 
The general agreement between our results and those in Ref.~\cite{ANGULO2003211} is not
surprising, as the low-energy scattering data in Ref.~\cite{ANGULO2003211} play a very significant role in both analyses. 
It can be inferred from Table~\ref{Scattering_length} that the uncertainties in the $s$-wave scattering lengths have been reduced  by a factor of 5-8 compared to the previous experimental measurement~\cite{ANGULO2003211}. 
This lower uncertainty may reduce the overall uncertainty in $S_{17}$(0), as discussed by Descouvemont~\cite{Descouvemont_2004} and Baye~\cite{Baye_2000}.
Using the potential model of Baye, the uncertainty in $S_{17}$(0) due to the average scattering length $\bar{a_0}$ can be calculated using Eq.~(20) from Ref.~\cite{Baye_2000}. Using this approach, the uncertainty in the average scattering length $\bar{a_0}$ deduced in this work using Eq.~(\ref{avg_scatt_length}) contributes the very small uncertainty of $\pm$0.03$\%$ to $S_{17}$(0), although it is not clear how this uncertainty impacts the extrapolation error on the $S_{17}$(0) value deduced from capture data. 
\par
Besides this measurement, there is only one $^7$Be+p elastic scattering measurement below 1~MeV. 
The measurements above this energy are not in agreement with each other. To better constrain the fits and the \textit{R}-matrix parameters,
more precise measurements are needed. Measurements below the 634-keV resonance are most important for constraining the scattering lengths. 
However, the data at higher energies are also important. Ideally, new scattering measurements would span a wide range of energy, from below the 634-keV resonance to well above 1~MeV. 
Transfer reactions could shed more light onto the structure of $^8$B.
\begin{acknowledgments}
 We are grateful to R. J. deBoer for his assistance with AZURE2. We thank D. R. Phillips for taking part in constructive discussions. 
This work was supported in part by the U.S. Department of Energy under Grants No.~DE-NA0002905, No.~DE-NA0003883, No.~DE-FG02-88ER40387, and DE-FG02-93ER40789 and 
Contract No. DE-AC05-00OR22725 (ORNL). D. W. B. acknowledges support from NSF under Grant No. PHY-1713857. B. D. acknowledges support from NSERC, Canada. 
The work of K. Y. C. was supported in part by a National Research Foundation of Korea (NRF) Grants No. NRF-2016R1A5A1013277 and No.~NRF-2013M7A1A1075764 funded by the Korean government (MEST).
This research used resources of the Holifield Radioactive Ion Beam Facility, which was a DOE Office of Science User Facility 
operated by the Oak Ridge National Laboratory. The authors are grateful to the staff of the HRIBF whose hard work made the experiment possible. 
\end{acknowledgments}

\bibliography{PRCreferences}

\begin{thebibliography}{36}%
\makeatletter
\providecommand \@ifxundefined [1]{%
 \@ifx{#1\undefined}
}%
\providecommand \@ifnum [1]{%
 \ifnum #1\expandafter \@firstoftwo
 \else \expandafter \@secondoftwo
 \fi
}%
\providecommand \@ifx [1]{%
 \ifx #1\expandafter \@firstoftwo
 \else \expandafter \@secondoftwo
 \fi
}%
\providecommand \natexlab [1]{#1}%
\providecommand \enquote  [1]{``#1''}%
\providecommand \bibnamefont  [1]{#1}%
\providecommand \bibfnamefont [1]{#1}%
\providecommand \citenamefont [1]{#1}%
\providecommand \href@noop [0]{\@secondoftwo}%
\providecommand \href [0]{\begingroup \@sanitize@url \@href}%
\providecommand \@href[1]{\@@startlink{#1}\@@href}%
\providecommand \@@href[1]{\endgroup#1\@@endlink}%
\providecommand \@sanitize@url [0]{\catcode `\\12\catcode `\$12\catcode
  `\&12\catcode `\#12\catcode `\^12\catcode `\_12\catcode `\%12\relax}%
\providecommand \@@startlink[1]{}%
\providecommand \@@endlink[0]{}%
\providecommand \url  [0]{\begingroup\@sanitize@url \@url }%
\providecommand \@url [1]{\endgroup\@href {#1}{\urlprefix }}%
\providecommand \urlprefix  [0]{URL }%
\providecommand \Eprint [0]{\href }%
\providecommand \doibase [0]{http://dx.doi.org/}%
\providecommand \selectlanguage [0]{\@gobble}%
\providecommand \bibinfo  [0]{\@secondoftwo}%
\providecommand \bibfield  [0]{\@secondoftwo}%
\providecommand \translation [1]{[#1]}%
\providecommand \BibitemOpen [0]{}%
\providecommand \bibitemStop [0]{}%
\providecommand \bibitemNoStop [0]{.\EOS\space}%
\providecommand \EOS [0]{\spacefactor3000\relax}%
\providecommand \BibitemShut  [1]{\csname bibitem#1\endcsname}%
\let\auto@bib@innerbib\@empty
\bibitem [{\citenamefont {Aharmim}\ \emph {et~al.}(2013)\citenamefont {Aharmim}
  \emph {et~al.}}]{SNO2013}%
  \BibitemOpen
  \bibfield  {author} {\bibinfo {author} {\bibfnamefont {B.}~\bibnamefont
  {Aharmim}} \emph {et~al.} (\bibinfo {collaboration} {SNO Collaboration}),\
  }\href {\doibase 10.1103/PhysRevC.88.025501} {\bibfield  {journal} {\bibinfo
  {journal} {Phys. Rev. C}\ }\textbf {\bibinfo {volume} {88}},\ \bibinfo
  {pages} {025501} (\bibinfo {year} {2013})}\BibitemShut {NoStop}%
\bibitem [{\citenamefont {Abe}\ \emph {et~al.}(2016)\citenamefont {Abe} \emph
  {et~al.}}]{SuperK}%
  \BibitemOpen
  \bibfield  {author} {\bibinfo {author} {\bibfnamefont {K.}~\bibnamefont
  {Abe}} \emph {et~al.} (\bibinfo {collaboration} {Super-Kamiokande
  Collaboration}),\ }\href {\doibase 10.1103/PhysRevD.94.052010} {\bibfield
  {journal} {\bibinfo  {journal} {Phys. Rev. D}\ }\textbf {\bibinfo {volume}
  {94}},\ \bibinfo {pages} {052010} (\bibinfo {year} {2016})}\BibitemShut
  {NoStop}%
\bibitem [{\citenamefont {Haxton}\ \emph {et~al.}(2013)\citenamefont {Haxton},
  \citenamefont {Hamish~Robertson},\ and\ \citenamefont {Serenelli}}]{Haxton}%
  \BibitemOpen
  \bibfield  {author} {\bibinfo {author} {\bibfnamefont {W.~C.}\ \bibnamefont
  {Haxton}}, \bibinfo {author} {\bibfnamefont {R.~G.}\ \bibnamefont
  {Hamish~Robertson}}, \ and\ \bibinfo {author} {\bibfnamefont {A.~M.}\
  \bibnamefont {Serenelli}},\ }\href {\doibase
  10.1146/annurev-astro-081811-125539} {\bibfield  {journal} {\bibinfo
  {journal} {Annual Review of Astronomy and Astrophysics}\ }\textbf {\bibinfo
  {volume} {51}},\ \bibinfo {pages} {21} (\bibinfo {year} {2013})},\ \Eprint
  {http://arxiv.org/abs/https://doi.org/10.1146/annurev-astro-081811-125539}
  {https://doi.org/10.1146/annurev-astro-081811-125539} \BibitemShut {NoStop}%
\bibitem [{\citenamefont {{Bahcall}}\ \emph {et~al.}(1969)\citenamefont
  {{Bahcall}}, \citenamefont {{Bahcall}},\ and\ \citenamefont
  {{Ulrich}}}]{Bahcall_1969}%
  \BibitemOpen
  \bibfield  {author} {\bibinfo {author} {\bibfnamefont {J.~N.}\ \bibnamefont
  {{Bahcall}}}, \bibinfo {author} {\bibfnamefont {N.~A.}\ \bibnamefont
  {{Bahcall}}}, \ and\ \bibinfo {author} {\bibfnamefont {R.~K.}\ \bibnamefont
  {{Ulrich}}},\ }\href {\doibase 10.1086/149989} {\bibfield  {journal}
  {\bibinfo  {journal} {\apj}\ }\textbf {\bibinfo {volume} {156}},\ \bibinfo
  {pages} {559} (\bibinfo {year} {1969})}\BibitemShut {NoStop}%
\bibitem [{\citenamefont {{Adelberger}}\ \emph {et~al.}(2011)\citenamefont
  {{Adelberger}}, \citenamefont {{Garc{\'{\i}}a}}, \citenamefont {{Robertson}},
  \citenamefont {{Snover}}, \citenamefont {{Balantekin}}, \citenamefont
  {{Heeger}}, \citenamefont {{Ramsey-Musolf}}, \citenamefont {{Bemmerer}},
  \citenamefont {{Junghans}}, \citenamefont {{Bertulani}}, \citenamefont
  {{Chen}}, \citenamefont {{Costantini}}, \citenamefont {{Prati}},
  \citenamefont {{Couder}}, \citenamefont {{Uberseder}}, \citenamefont
  {{Wiescher}}, \citenamefont {{Cyburt}}, \citenamefont {{Davids}},
  \citenamefont {{Freedman}}, \citenamefont {{Gai}}, \citenamefont {{Gazit}},
  \citenamefont {{Gialanella}}, \citenamefont {{Imbriani}}, \citenamefont
  {{Greife}}, \citenamefont {{Hass}}, \citenamefont {{Haxton}}, \citenamefont
  {{Itahashi}}, \citenamefont {{Kubodera}}, \citenamefont {{Langanke}},
  \citenamefont {{Leitner}}, \citenamefont {{Leitner}}, \citenamefont
  {{Vetter}}, \citenamefont {{Winslow}}, \citenamefont {{Marcucci}},
  \citenamefont {{Motobayashi}}, \citenamefont {{Mukhamedzhanov}},
  \citenamefont {{Tribble}}, \citenamefont {{Nollett}}, \citenamefont
  {{Nunes}}, \citenamefont {{Park}}, \citenamefont {{Parker}}, \citenamefont
  {{Schiavilla}}, \citenamefont {{Simpson}}, \citenamefont {{Spitaleri}},
  \citenamefont {{Strieder}}, \citenamefont {{Trautvetter}}, \citenamefont
  {{Suemmerer}},\ and\ \citenamefont {{Typel}}}]{adelberger11}%
  \BibitemOpen
  \bibfield  {author} {\bibinfo {author} {\bibfnamefont {E.~G.}\ \bibnamefont
  {{Adelberger}}}, \bibinfo {author} {\bibfnamefont {A.}~\bibnamefont
  {{Garc{\'{\i}}a}}}, \bibinfo {author} {\bibfnamefont {R.~G.~H.}\ \bibnamefont
  {{Robertson}}}, \bibinfo {author} {\bibfnamefont {K.~A.}\ \bibnamefont
  {{Snover}}}, \bibinfo {author} {\bibfnamefont {A.~B.}\ \bibnamefont
  {{Balantekin}}}, \bibinfo {author} {\bibfnamefont {K.}~\bibnamefont
  {{Heeger}}}, \bibinfo {author} {\bibfnamefont {M.~J.}\ \bibnamefont
  {{Ramsey-Musolf}}}, \bibinfo {author} {\bibfnamefont {D.}~\bibnamefont
  {{Bemmerer}}}, \bibinfo {author} {\bibfnamefont {A.}~\bibnamefont
  {{Junghans}}}, \bibinfo {author} {\bibfnamefont {C.~A.}\ \bibnamefont
  {{Bertulani}}}, \bibinfo {author} {\bibfnamefont {J.-W.}\ \bibnamefont
  {{Chen}}}, \bibinfo {author} {\bibfnamefont {H.}~\bibnamefont
  {{Costantini}}}, \bibinfo {author} {\bibfnamefont {P.}~\bibnamefont
  {{Prati}}}, \bibinfo {author} {\bibfnamefont {M.}~\bibnamefont {{Couder}}},
  \bibinfo {author} {\bibfnamefont {E.}~\bibnamefont {{Uberseder}}}, \bibinfo
  {author} {\bibfnamefont {M.}~\bibnamefont {{Wiescher}}}, \bibinfo {author}
  {\bibfnamefont {R.}~\bibnamefont {{Cyburt}}}, \bibinfo {author}
  {\bibfnamefont {B.}~\bibnamefont {{Davids}}}, \bibinfo {author}
  {\bibfnamefont {S.~J.}\ \bibnamefont {{Freedman}}}, \bibinfo {author}
  {\bibfnamefont {M.}~\bibnamefont {{Gai}}}, \bibinfo {author} {\bibfnamefont
  {D.}~\bibnamefont {{Gazit}}}, \bibinfo {author} {\bibfnamefont
  {L.}~\bibnamefont {{Gialanella}}}, \bibinfo {author} {\bibfnamefont
  {G.}~\bibnamefont {{Imbriani}}}, \bibinfo {author} {\bibfnamefont
  {U.}~\bibnamefont {{Greife}}}, \bibinfo {author} {\bibfnamefont
  {M.}~\bibnamefont {{Hass}}}, \bibinfo {author} {\bibfnamefont {W.~C.}\
  \bibnamefont {{Haxton}}}, \bibinfo {author} {\bibfnamefont {T.}~\bibnamefont
  {{Itahashi}}}, \bibinfo {author} {\bibfnamefont {K.}~\bibnamefont
  {{Kubodera}}}, \bibinfo {author} {\bibfnamefont {K.}~\bibnamefont
  {{Langanke}}}, \bibinfo {author} {\bibfnamefont {D.}~\bibnamefont
  {{Leitner}}}, \bibinfo {author} {\bibfnamefont {M.}~\bibnamefont
  {{Leitner}}}, \bibinfo {author} {\bibfnamefont {P.}~\bibnamefont {{Vetter}}},
  \bibinfo {author} {\bibfnamefont {L.}~\bibnamefont {{Winslow}}}, \bibinfo
  {author} {\bibfnamefont {L.~E.}\ \bibnamefont {{Marcucci}}}, \bibinfo
  {author} {\bibfnamefont {T.}~\bibnamefont {{Motobayashi}}}, \bibinfo {author}
  {\bibfnamefont {A.}~\bibnamefont {{Mukhamedzhanov}}}, \bibinfo {author}
  {\bibfnamefont {R.~E.}\ \bibnamefont {{Tribble}}}, \bibinfo {author}
  {\bibfnamefont {K.~M.}\ \bibnamefont {{Nollett}}}, \bibinfo {author}
  {\bibfnamefont {F.~M.}\ \bibnamefont {{Nunes}}}, \bibinfo {author}
  {\bibfnamefont {T.-S.}\ \bibnamefont {{Park}}}, \bibinfo {author}
  {\bibfnamefont {P.~D.}\ \bibnamefont {{Parker}}}, \bibinfo {author}
  {\bibfnamefont {R.}~\bibnamefont {{Schiavilla}}}, \bibinfo {author}
  {\bibfnamefont {E.~C.}\ \bibnamefont {{Simpson}}}, \bibinfo {author}
  {\bibfnamefont {C.}~\bibnamefont {{Spitaleri}}}, \bibinfo {author}
  {\bibfnamefont {F.}~\bibnamefont {{Strieder}}}, \bibinfo {author}
  {\bibfnamefont {H.-P.}\ \bibnamefont {{Trautvetter}}}, \bibinfo {author}
  {\bibfnamefont {K.}~\bibnamefont {{Suemmerer}}}, \ and\ \bibinfo {author}
  {\bibfnamefont {S.}~\bibnamefont {{Typel}}},\ }\href {\doibase
  10.1103/RevModPhys.83.195} {\bibfield  {journal} {\bibinfo  {journal}
  {Reviews of Modern Physics}\ }\textbf {\bibinfo {volume} {83}},\ \bibinfo
  {pages} {195} (\bibinfo {year} {2011})}\BibitemShut {NoStop}%
\bibitem [{\citenamefont {{Brune}}\ and\ \citenamefont
  {{Davids}}(2015)}]{brune15}%
  \BibitemOpen
  \bibfield  {author} {\bibinfo {author} {\bibfnamefont {C.~R.}\ \bibnamefont
  {{Brune}}}\ and\ \bibinfo {author} {\bibfnamefont {B.}~\bibnamefont
  {{Davids}}},\ }\href {\doibase 10.1146/annurev-nucl-102014-022027} {\bibfield
   {journal} {\bibinfo  {journal} {Annual Review of Nuclear and Particle
  Science}\ }\textbf {\bibinfo {volume} {65}},\ \bibinfo {pages} {87} (\bibinfo
  {year} {2015})}\BibitemShut {NoStop}%
\bibitem [{\citenamefont {Descouvemont}(2004)}]{Descouvemont_2004}%
  \BibitemOpen
  \bibfield  {author} {\bibinfo {author} {\bibfnamefont {P.}~\bibnamefont
  {Descouvemont}},\ }\href {\doibase 10.1103/PhysRevC.70.065802} {\bibfield
  {journal} {\bibinfo  {journal} {Phys. Rev. C}\ }\textbf {\bibinfo {volume}
  {70}},\ \bibinfo {pages} {065802} (\bibinfo {year} {2004})}\BibitemShut
  {NoStop}%
\bibitem [{\citenamefont {Zhang}\ \emph {et~al.}(2014)\citenamefont {Zhang},
  \citenamefont {Nollett},\ and\ \citenamefont {Phillips}}]{Zhang_2014}%
  \BibitemOpen
  \bibfield  {author} {\bibinfo {author} {\bibfnamefont {X.}~\bibnamefont
  {Zhang}}, \bibinfo {author} {\bibfnamefont {K.~M.}\ \bibnamefont {Nollett}},
  \ and\ \bibinfo {author} {\bibfnamefont {D.~R.}\ \bibnamefont {Phillips}},\
  }\href {\doibase 10.1103/PhysRevC.89.051602} {\bibfield  {journal} {\bibinfo
  {journal} {Phys. Rev. C}\ }\textbf {\bibinfo {volume} {89}},\ \bibinfo
  {pages} {051602} (\bibinfo {year} {2014})}\BibitemShut {NoStop}%
\bibitem [{\citenamefont {Baye}(2000)}]{Baye_2000}%
  \BibitemOpen
  \bibfield  {author} {\bibinfo {author} {\bibfnamefont {D.}~\bibnamefont
  {Baye}},\ }\href {\doibase 10.1103/PhysRevC.62.065803} {\bibfield  {journal}
  {\bibinfo  {journal} {Phys. Rev. C}\ }\textbf {\bibinfo {volume} {62}},\
  \bibinfo {pages} {065803} (\bibinfo {year} {2000})}\BibitemShut {NoStop}%
\bibitem [{\citenamefont {Zhang}\ \emph {et~al.}(2015)\citenamefont {Zhang},
  \citenamefont {Nollett},\ and\ \citenamefont {Phillips}}]{Zhang2015}%
  \BibitemOpen
  \bibfield  {author} {\bibinfo {author} {\bibfnamefont {X.}~\bibnamefont
  {Zhang}}, \bibinfo {author} {\bibfnamefont {K.~M.}\ \bibnamefont {Nollett}},
  \ and\ \bibinfo {author} {\bibfnamefont {D.}~\bibnamefont {Phillips}},\
  }\href {\doibase https://doi.org/10.1016/j.physletb.2015.11.005} {\bibfield
  {journal} {\bibinfo  {journal} {Physics Letters B}\ }\textbf {\bibinfo
  {volume} {751}},\ \bibinfo {pages} {535 } (\bibinfo {year}
  {2015})}\BibitemShut {NoStop}%
\bibitem [{\citenamefont {Angulo}\ \emph {et~al.}(2003)\citenamefont {Angulo},
  \citenamefont {Azzouz}, \citenamefont {Descouvemont}, \citenamefont
  {Tabacaru}, \citenamefont {Baye}, \citenamefont {Cogneau}, \citenamefont
  {Couder}, \citenamefont {Davinson}, \citenamefont {Pietro}, \citenamefont
  {Figuera}, \citenamefont {Gaelens}, \citenamefont {Leleux}, \citenamefont
  {Loiselet}, \citenamefont {Ninane}, \citenamefont {de~Oliveira~Santos},
  \citenamefont {Pizzone}, \citenamefont {Ryckewaert}, \citenamefont
  {de Séréville},\ and\ \citenamefont {Vanderbist}}]{ANGULO2003211}%
  \BibitemOpen
  \bibfield  {author} {\bibinfo {author} {\bibfnamefont {C.}~\bibnamefont
  {Angulo}}, \bibinfo {author} {\bibfnamefont {M.}~\bibnamefont {Azzouz}},
  \bibinfo {author} {\bibfnamefont {P.}~\bibnamefont {Descouvemont}}, \bibinfo
  {author} {\bibfnamefont {G.}~\bibnamefont {Tabacaru}}, \bibinfo {author}
  {\bibfnamefont {D.}~\bibnamefont {Baye}}, \bibinfo {author} {\bibfnamefont
  {M.}~\bibnamefont {Cogneau}}, \bibinfo {author} {\bibfnamefont
  {M.}~\bibnamefont {Couder}}, \bibinfo {author} {\bibfnamefont
  {T.}~\bibnamefont {Davinson}}, \bibinfo {author} {\bibfnamefont {A.~D.}\
  \bibnamefont {Pietro}}, \bibinfo {author} {\bibfnamefont {P.}~\bibnamefont
  {Figuera}}, \bibinfo {author} {\bibfnamefont {M.}~\bibnamefont {Gaelens}},
  \bibinfo {author} {\bibfnamefont {P.}~\bibnamefont {Leleux}}, \bibinfo
  {author} {\bibfnamefont {M.}~\bibnamefont {Loiselet}}, \bibinfo {author}
  {\bibfnamefont {A.}~\bibnamefont {Ninane}}, \bibinfo {author} {\bibfnamefont
  {F.}~\bibnamefont {de~Oliveira~Santos}}, \bibinfo {author} {\bibfnamefont
  {R.}~\bibnamefont {Pizzone}}, \bibinfo {author} {\bibfnamefont
  {G.}~\bibnamefont {Ryckewaert}}, \bibinfo {author} {\bibfnamefont
  {N.}~\bibnamefont {de Séréville}}, \ and\ \bibinfo {author} {\bibfnamefont
  {F.}~\bibnamefont {Vanderbist}},\ }\href {\doibase
  http://dx.doi.org/10.1016/S0375-9474(02)01584-1} {\bibfield  {journal}
  {\bibinfo  {journal} {Nuclear Physics A}\ }\textbf {\bibinfo {volume}
  {716}},\ \bibinfo {pages} {211 } (\bibinfo {year} {2003})}\BibitemShut
  {NoStop}%
\bibitem [{\citenamefont {Navratil}\ \emph {et~al.}(2011)\citenamefont
  {Navratil}, \citenamefont {Roth},\ and\ \citenamefont
  {Quaglioni}}]{NAVRATIL2011}%
  \BibitemOpen
  \bibfield  {author} {\bibinfo {author} {\bibfnamefont {P.}~\bibnamefont
  {Navratil}}, \bibinfo {author} {\bibfnamefont {R.}~\bibnamefont {Roth}}, \
  and\ \bibinfo {author} {\bibfnamefont {S.}~\bibnamefont {Quaglioni}},\ }\href
  {\doibase https://doi.org/10.1016/j.physletb.2011.09.079} {\bibfield
  {journal} {\bibinfo  {journal} {Physics Letters B}\ }\textbf {\bibinfo
  {volume} {704}},\ \bibinfo {pages} {379 } (\bibinfo {year}
  {2011})}\BibitemShut {NoStop}%
\bibitem [{\citenamefont {Tilley}\ \emph {et~al.}(2004)\citenamefont {Tilley},
  \citenamefont {Kelley}, \citenamefont {Godwin}, \citenamefont {Millener},
  \citenamefont {Purcell}, \citenamefont {Sheu},\ and\ \citenamefont
  {Weller}}]{tilley04}%
  \BibitemOpen
  \bibfield  {author} {\bibinfo {author} {\bibfnamefont {D.}~\bibnamefont
  {Tilley}}, \bibinfo {author} {\bibfnamefont {J.}~\bibnamefont {Kelley}},
  \bibinfo {author} {\bibfnamefont {J.}~\bibnamefont {Godwin}}, \bibinfo
  {author} {\bibfnamefont {D.}~\bibnamefont {Millener}}, \bibinfo {author}
  {\bibfnamefont {J.}~\bibnamefont {Purcell}}, \bibinfo {author} {\bibfnamefont
  {C.}~\bibnamefont {Sheu}}, \ and\ \bibinfo {author} {\bibfnamefont
  {H.}~\bibnamefont {Weller}},\ }\href {\doibase
  https://doi.org/10.1016/j.nuclphysa.2004.09.059} {\bibfield  {journal}
  {\bibinfo  {journal} {Nuclear Physics A}\ }\textbf {\bibinfo {volume}
  {745}},\ \bibinfo {pages} {155 } (\bibinfo {year} {2004})}\BibitemShut
  {NoStop}%
\bibitem [{\citenamefont {Gol'dberg}\ \emph {et~al.}(1998)\citenamefont
  {Gol'dberg}, \citenamefont {Rogachev}, \citenamefont {Golovkov},
  \citenamefont {Dukhanov}, \citenamefont {Serikov},\ and\ \citenamefont
  {Timofeev}}]{Goldberg1998}%
  \BibitemOpen
  \bibfield  {author} {\bibinfo {author} {\bibfnamefont {V.~Z.}\ \bibnamefont
  {Gol'dberg}}, \bibinfo {author} {\bibfnamefont {G.~V.}\ \bibnamefont
  {Rogachev}}, \bibinfo {author} {\bibfnamefont {M.~S.}\ \bibnamefont
  {Golovkov}}, \bibinfo {author} {\bibfnamefont {V.~I.}\ \bibnamefont
  {Dukhanov}}, \bibinfo {author} {\bibfnamefont {I.~N.}\ \bibnamefont
  {Serikov}}, \ and\ \bibinfo {author} {\bibfnamefont {V.~A.}\ \bibnamefont
  {Timofeev}},\ }\href {\doibase 10.1134/1.567784} {\bibfield  {journal}
  {\bibinfo  {journal} {Journal of Experimental and Theoretical Physics
  Letters}\ }\textbf {\bibinfo {volume} {67}},\ \bibinfo {pages} {1013}
  (\bibinfo {year} {1998})}\BibitemShut {NoStop}%
\bibitem [{\citenamefont {Rogachev}\ \emph {et~al.}(2001)\citenamefont
  {Rogachev}, \citenamefont {Kolata}, \citenamefont {Becchetti}, \citenamefont
  {DeYoung}, \citenamefont {Hencheck}, \citenamefont {Helland}, \citenamefont
  {Hinnefeld}, \citenamefont {Hughey}, \citenamefont {Jolivette}, \citenamefont
  {Kiessel}, \citenamefont {Lee}, \citenamefont {Lee}, \citenamefont
  {O'Donnell}, \citenamefont {Peaslee}, \citenamefont {Peterson}, \citenamefont
  {Roberts}, \citenamefont {Santi},\ and\ \citenamefont
  {Shaheen}}]{Rogachev_2001}%
  \BibitemOpen
  \bibfield  {author} {\bibinfo {author} {\bibfnamefont {G.~V.}\ \bibnamefont
  {Rogachev}}, \bibinfo {author} {\bibfnamefont {J.~J.}\ \bibnamefont
  {Kolata}}, \bibinfo {author} {\bibfnamefont {F.~D.}\ \bibnamefont
  {Becchetti}}, \bibinfo {author} {\bibfnamefont {P.~A.}\ \bibnamefont
  {DeYoung}}, \bibinfo {author} {\bibfnamefont {M.}~\bibnamefont {Hencheck}},
  \bibinfo {author} {\bibfnamefont {K.}~\bibnamefont {Helland}}, \bibinfo
  {author} {\bibfnamefont {J.~D.}\ \bibnamefont {Hinnefeld}}, \bibinfo {author}
  {\bibfnamefont {B.}~\bibnamefont {Hughey}}, \bibinfo {author} {\bibfnamefont
  {P.~L.}\ \bibnamefont {Jolivette}}, \bibinfo {author} {\bibfnamefont {L.~M.}\
  \bibnamefont {Kiessel}}, \bibinfo {author} {\bibfnamefont {H.-Y.}\
  \bibnamefont {Lee}}, \bibinfo {author} {\bibfnamefont {M.~Y.}\ \bibnamefont
  {Lee}}, \bibinfo {author} {\bibfnamefont {T.~W.}\ \bibnamefont {O'Donnell}},
  \bibinfo {author} {\bibfnamefont {G.~F.}\ \bibnamefont {Peaslee}}, \bibinfo
  {author} {\bibfnamefont {D.}~\bibnamefont {Peterson}}, \bibinfo {author}
  {\bibfnamefont {D.~A.}\ \bibnamefont {Roberts}}, \bibinfo {author}
  {\bibfnamefont {P.}~\bibnamefont {Santi}}, \ and\ \bibinfo {author}
  {\bibfnamefont {S.~A.}\ \bibnamefont {Shaheen}},\ }\href {\doibase
  10.1103/PhysRevC.64.061601} {\bibfield  {journal} {\bibinfo  {journal} {Phys.
  Rev. C}\ }\textbf {\bibinfo {volume} {64}},\ \bibinfo {pages} {061601}
  (\bibinfo {year} {2001})}\BibitemShut {NoStop}%
\bibitem [{\citenamefont {Yamaguchi}\ \emph {et~al.}(2009)\citenamefont
  {Yamaguchi}, \citenamefont {Wakabayashi}, \citenamefont {Kubono},
  \citenamefont {Amadio}, \citenamefont {Fujikawa}, \citenamefont {Teranishi},
  \citenamefont {Saito}, \citenamefont {He}, \citenamefont {Nishimura},
  \citenamefont {Togano}, \citenamefont {Kwon}, \citenamefont {Niikura},
  \citenamefont {Iwasa}, \citenamefont {Inafuku},\ and\ \citenamefont
  {Khiem}}]{Yamaguchi2009}%
  \BibitemOpen
  \bibfield  {author} {\bibinfo {author} {\bibfnamefont {H.}~\bibnamefont
  {Yamaguchi}}, \bibinfo {author} {\bibfnamefont {Y.}~\bibnamefont
  {Wakabayashi}}, \bibinfo {author} {\bibfnamefont {S.}~\bibnamefont {Kubono}},
  \bibinfo {author} {\bibfnamefont {G.}~\bibnamefont {Amadio}}, \bibinfo
  {author} {\bibfnamefont {H.}~\bibnamefont {Fujikawa}}, \bibinfo {author}
  {\bibfnamefont {T.}~\bibnamefont {Teranishi}}, \bibinfo {author}
  {\bibfnamefont {A.}~\bibnamefont {Saito}}, \bibinfo {author} {\bibfnamefont
  {J.}~\bibnamefont {He}}, \bibinfo {author} {\bibfnamefont {S.}~\bibnamefont
  {Nishimura}}, \bibinfo {author} {\bibfnamefont {Y.}~\bibnamefont {Togano}},
  \bibinfo {author} {\bibfnamefont {Y.}~\bibnamefont {Kwon}}, \bibinfo {author}
  {\bibfnamefont {M.}~\bibnamefont {Niikura}}, \bibinfo {author} {\bibfnamefont
  {N.}~\bibnamefont {Iwasa}}, \bibinfo {author} {\bibfnamefont
  {K.}~\bibnamefont {Inafuku}}, \ and\ \bibinfo {author} {\bibfnamefont
  {L.}~\bibnamefont {Khiem}},\ }\href {\doibase
  https://doi.org/10.1016/j.physletb.2009.01.033} {\bibfield  {journal}
  {\bibinfo  {journal} {Physics Letters B}\ }\textbf {\bibinfo {volume}
  {672}},\ \bibinfo {pages} {230 } (\bibinfo {year} {2009})}\BibitemShut
  {NoStop}%
\bibitem [{\citenamefont {Mitchell}\ \emph {et~al.}(2013)\citenamefont
  {Mitchell}, \citenamefont {Rogachev}, \citenamefont {Johnson}, \citenamefont
  {Baby}, \citenamefont {Kemper}, \citenamefont {Moro}, \citenamefont
  {Peplowski}, \citenamefont {Volya},\ and\ \citenamefont
  {Wiedenh\"over}}]{Mitchell_2013}%
  \BibitemOpen
  \bibfield  {author} {\bibinfo {author} {\bibfnamefont {J.~P.}\ \bibnamefont
  {Mitchell}}, \bibinfo {author} {\bibfnamefont {G.~V.}\ \bibnamefont
  {Rogachev}}, \bibinfo {author} {\bibfnamefont {E.~D.}\ \bibnamefont
  {Johnson}}, \bibinfo {author} {\bibfnamefont {L.~T.}\ \bibnamefont {Baby}},
  \bibinfo {author} {\bibfnamefont {K.~W.}\ \bibnamefont {Kemper}}, \bibinfo
  {author} {\bibfnamefont {A.~M.}\ \bibnamefont {Moro}}, \bibinfo {author}
  {\bibfnamefont {P.}~\bibnamefont {Peplowski}}, \bibinfo {author}
  {\bibfnamefont {A.~S.}\ \bibnamefont {Volya}}, \ and\ \bibinfo {author}
  {\bibfnamefont {I.}~\bibnamefont {Wiedenh\"over}},\ }\href {\doibase
  10.1103/PhysRevC.87.054617} {\bibfield  {journal} {\bibinfo  {journal} {Phys.
  Rev. C}\ }\textbf {\bibinfo {volume} {87}},\ \bibinfo {pages} {054617}
  (\bibinfo {year} {2013})}\BibitemShut {NoStop}%
\bibitem [{\citenamefont {Lane}\ and\ \citenamefont
  {Thomas}(1958)}]{Lane_Thomas}%
  \BibitemOpen
  \bibfield  {author} {\bibinfo {author} {\bibfnamefont {A.~M.}\ \bibnamefont
  {Lane}}\ and\ \bibinfo {author} {\bibfnamefont {R.~G.}\ \bibnamefont
  {Thomas}},\ }\href {\doibase 10.1103/RevModPhys.30.257} {\bibfield  {journal}
  {\bibinfo  {journal} {Rev. Mod. Phys.}\ }\textbf {\bibinfo {volume} {30}},\
  \bibinfo {pages} {257} (\bibinfo {year} {1958})}\BibitemShut {NoStop}%
\bibitem [{\citenamefont {Beene}\ \emph {et~al.}(2011)\citenamefont {Beene},
  \citenamefont {Bardayan}, \citenamefont {Uribarri}, \citenamefont {Gross},
  \citenamefont {Jones}, \citenamefont {Liang}, \citenamefont {Nazarewicz},
  \citenamefont {Stracener}, \citenamefont {Tatum},\ and\ \citenamefont
  {Varner}}]{HRIBF_2011}%
  \BibitemOpen
  \bibfield  {author} {\bibinfo {author} {\bibfnamefont {J.~R.}\ \bibnamefont
  {Beene}}, \bibinfo {author} {\bibfnamefont {D.~W.}\ \bibnamefont {Bardayan}},
  \bibinfo {author} {\bibfnamefont {A.~G.}\ \bibnamefont {Uribarri}}, \bibinfo
  {author} {\bibfnamefont {C.~J.}\ \bibnamefont {Gross}}, \bibinfo {author}
  {\bibfnamefont {K.~L.}\ \bibnamefont {Jones}}, \bibinfo {author}
  {\bibfnamefont {J.~F.}\ \bibnamefont {Liang}}, \bibinfo {author}
  {\bibfnamefont {W.}~\bibnamefont {Nazarewicz}}, \bibinfo {author}
  {\bibfnamefont {D.~W.}\ \bibnamefont {Stracener}}, \bibinfo {author}
  {\bibfnamefont {B.~A.}\ \bibnamefont {Tatum}}, \ and\ \bibinfo {author}
  {\bibfnamefont {R.~L.}\ \bibnamefont {Varner}},\ }\href
  {http://stacks.iop.org/0954-3899/38/i=2/a=024002} {\bibfield  {journal}
  {\bibinfo  {journal} {Journal of Physics G: Nuclear and Particle Physics}\
  }\textbf {\bibinfo {volume} {38}},\ \bibinfo {pages} {024002} (\bibinfo
  {year} {2011})}\BibitemShut {NoStop}%
\bibitem [{\citenamefont {Fitzgerald}(2005)}]{Fitzgerald}%
  \BibitemOpen
  \bibfield  {author} {\bibinfo {author} {\bibfnamefont {R.~P.}\ \bibnamefont
  {Fitzgerald}},\ }\href@noop {} {\bibinfo {type} {{Ph.D.} dissertation}},\
  \bibinfo  {school} {University of North Carolina, Chapel Hill} (\bibinfo
  {year} {2005})\BibitemShut {NoStop}%
\bibitem [{\citenamefont {Gialanella}\ \emph {et~al.}(2002)\citenamefont
  {Gialanella}, \citenamefont {Greife}, \citenamefont {Cesare}, \citenamefont
  {D'Onofrio}, \citenamefont {Romano}, \citenamefont {Campajola}, \citenamefont
  {Formicola}, \citenamefont {Fulop}, \citenamefont {Gy\"{u}rky}, \citenamefont
  {Imbriani}, \citenamefont {Lubritto}, \citenamefont {Ordine}, \citenamefont
  {Roca}, \citenamefont {Rogalla}, \citenamefont {Rolfs}, \citenamefont
  {Russo}, \citenamefont {Sabbarese}, \citenamefont {Somorjai}, \citenamefont
  {Strieder}, \citenamefont {Terrasi},\ and\ \citenamefont
  {Trautvetter}}]{chemical_extraction}%
  \BibitemOpen
  \bibfield  {author} {\bibinfo {author} {\bibfnamefont {L.}~\bibnamefont
  {Gialanella}}, \bibinfo {author} {\bibfnamefont {U.}~\bibnamefont {Greife}},
  \bibinfo {author} {\bibfnamefont {N.~D.}\ \bibnamefont {Cesare}}, \bibinfo
  {author} {\bibfnamefont {A.}~\bibnamefont {D'Onofrio}}, \bibinfo {author}
  {\bibfnamefont {M.}~\bibnamefont {Romano}}, \bibinfo {author} {\bibfnamefont
  {L.}~\bibnamefont {Campajola}}, \bibinfo {author} {\bibfnamefont
  {A.}~\bibnamefont {Formicola}}, \bibinfo {author} {\bibfnamefont
  {Z.}~\bibnamefont {Fulop}}, \bibinfo {author} {\bibfnamefont
  {G.}~\bibnamefont {Gy\"{u}rky}}, \bibinfo {author} {\bibfnamefont
  {G.}~\bibnamefont {Imbriani}}, \bibinfo {author} {\bibfnamefont
  {C.}~\bibnamefont {Lubritto}}, \bibinfo {author} {\bibfnamefont
  {A.}~\bibnamefont {Ordine}}, \bibinfo {author} {\bibfnamefont
  {V.}~\bibnamefont {Roca}}, \bibinfo {author} {\bibfnamefont {D.}~\bibnamefont
  {Rogalla}}, \bibinfo {author} {\bibfnamefont {C.}~\bibnamefont {Rolfs}},
  \bibinfo {author} {\bibfnamefont {M.}~\bibnamefont {Russo}}, \bibinfo
  {author} {\bibfnamefont {C.}~\bibnamefont {Sabbarese}}, \bibinfo {author}
  {\bibfnamefont {E.}~\bibnamefont {Somorjai}}, \bibinfo {author}
  {\bibfnamefont {F.}~\bibnamefont {Strieder}}, \bibinfo {author}
  {\bibfnamefont {F.}~\bibnamefont {Terrasi}}, \ and\ \bibinfo {author}
  {\bibfnamefont {H.}~\bibnamefont {Trautvetter}},\ }\href {\doibase
  https://doi.org/10.1016/S0168-583X(02)01386-1} {\bibfield  {journal}
  {\bibinfo  {journal} {Nuclear Instruments and Methods in Physics Research
  Section B: Beam Interactions with Materials and Atoms}\ }\textbf {\bibinfo
  {volume} {197}},\ \bibinfo {pages} {150 } (\bibinfo {year}
  {2002})}\BibitemShut {NoStop}%
\bibitem [{\citenamefont {Bardayan}\ \emph {et~al.}(2000)\citenamefont
  {Bardayan}, \citenamefont {Blackmon}, \citenamefont {Brune}, \citenamefont
  {Champagne}, \citenamefont {Chen}, \citenamefont {Cox}, \citenamefont
  {Davinson}, \citenamefont {Hansper}, \citenamefont {Hofstee}, \citenamefont
  {Johnson}, \citenamefont {Kozub}, \citenamefont {Ma}, \citenamefont {Parker},
  \citenamefont {Pierce}, \citenamefont {Rabban}, \citenamefont {Shotter},
  \citenamefont {Smith}, \citenamefont {Swartz}, \citenamefont {Visser},\ and\
  \citenamefont {Woods}}]{SIDAR_2000}%
  \BibitemOpen
  \bibfield  {author} {\bibinfo {author} {\bibfnamefont {D.~W.}\ \bibnamefont
  {Bardayan}}, \bibinfo {author} {\bibfnamefont {J.~C.}\ \bibnamefont
  {Blackmon}}, \bibinfo {author} {\bibfnamefont {C.~R.}\ \bibnamefont {Brune}},
  \bibinfo {author} {\bibfnamefont {A.~E.}\ \bibnamefont {Champagne}}, \bibinfo
  {author} {\bibfnamefont {A.~A.}\ \bibnamefont {Chen}}, \bibinfo {author}
  {\bibfnamefont {J.~M.}\ \bibnamefont {Cox}}, \bibinfo {author} {\bibfnamefont
  {T.}~\bibnamefont {Davinson}}, \bibinfo {author} {\bibfnamefont {V.~Y.}\
  \bibnamefont {Hansper}}, \bibinfo {author} {\bibfnamefont {M.~A.}\
  \bibnamefont {Hofstee}}, \bibinfo {author} {\bibfnamefont {B.~A.}\
  \bibnamefont {Johnson}}, \bibinfo {author} {\bibfnamefont {R.~L.}\
  \bibnamefont {Kozub}}, \bibinfo {author} {\bibfnamefont {Z.}~\bibnamefont
  {Ma}}, \bibinfo {author} {\bibfnamefont {P.~D.}\ \bibnamefont {Parker}},
  \bibinfo {author} {\bibfnamefont {D.~E.}\ \bibnamefont {Pierce}}, \bibinfo
  {author} {\bibfnamefont {M.~T.}\ \bibnamefont {Rabban}}, \bibinfo {author}
  {\bibfnamefont {A.~C.}\ \bibnamefont {Shotter}}, \bibinfo {author}
  {\bibfnamefont {M.~S.}\ \bibnamefont {Smith}}, \bibinfo {author}
  {\bibfnamefont {K.~B.}\ \bibnamefont {Swartz}}, \bibinfo {author}
  {\bibfnamefont {D.~W.}\ \bibnamefont {Visser}}, \ and\ \bibinfo {author}
  {\bibfnamefont {P.~J.}\ \bibnamefont {Woods}},\ }\href {\doibase
  10.1103/PhysRevC.62.055804} {\bibfield  {journal} {\bibinfo  {journal} {Phys.
  Rev. C}\ }\textbf {\bibinfo {volume} {62}},\ \bibinfo {pages} {055804}
  (\bibinfo {year} {2000})}\BibitemShut {NoStop}%
\bibitem [{\citenamefont {Livesay}(2006)}]{Livesay}%
  \BibitemOpen
  \bibfield  {author} {\bibinfo {author} {\bibfnamefont {R.}~\bibnamefont
  {Livesay}},\ }\href@noop {} {\bibinfo {type} {{Ph.D.} dissertation}},\
  \bibinfo  {school} {Colorado School of Mines} (\bibinfo {year}
  {2006})\BibitemShut {NoStop}%
\bibitem [{\citenamefont {Ziegler}\ and\ \citenamefont {Biersack}()}]{SRIM}%
  \BibitemOpen
  \bibfield  {author} {\bibinfo {author} {\bibfnamefont {J.}~\bibnamefont
  {Ziegler}}\ and\ \bibinfo {author} {\bibfnamefont {J.}~\bibnamefont
  {Biersack}},\ }\href@noop {} {\enquote {\bibinfo {title} {{Transport of Ions
  in Matter, TRIM program Version 95.06, SRIM 2003}},}\ }\bibinfo
  {howpublished} {\url{https://www.srim.org}}\BibitemShut {NoStop}%
\bibitem [{\citenamefont {Brune}\ and\ \citenamefont
  {Sayre}(2013)}]{BRUNE2013}%
  \BibitemOpen
  \bibfield  {author} {\bibinfo {author} {\bibfnamefont {C.~R.}\ \bibnamefont
  {Brune}}\ and\ \bibinfo {author} {\bibfnamefont {D.~B.}\ \bibnamefont
  {Sayre}},\ }\href {\doibase https://doi.org/10.1016/j.nima.2012.09.023}
  {\bibfield  {journal} {\bibinfo  {journal} {Nuclear Instruments and Methods
  in Physics Research Section A: Accelerators, Spectrometers, Detectors and
  Associated Equipment}\ }\textbf {\bibinfo {volume} {698}},\ \bibinfo {pages}
  {49 } (\bibinfo {year} {2013})}\BibitemShut {NoStop}%
\bibitem [{\citenamefont {Poling}\ \emph {et~al.}(1976)\citenamefont {Poling},
  \citenamefont {Norbeck},\ and\ \citenamefont {Carlson}}]{Poling_1976}%
  \BibitemOpen
  \bibfield  {author} {\bibinfo {author} {\bibfnamefont {J.~E.}\ \bibnamefont
  {Poling}}, \bibinfo {author} {\bibfnamefont {E.}~\bibnamefont {Norbeck}}, \
  and\ \bibinfo {author} {\bibfnamefont {R.~R.}\ \bibnamefont {Carlson}},\
  }\href {\doibase 10.1103/PhysRevC.13.648} {\bibfield  {journal} {\bibinfo
  {journal} {Phys. Rev. C}\ }\textbf {\bibinfo {volume} {13}},\ \bibinfo
  {pages} {648} (\bibinfo {year} {1976})}\BibitemShut {NoStop}%
\bibitem [{\citenamefont {Kunz}()}]{DWUCK5}%
  \BibitemOpen
  \bibfield  {author} {\bibinfo {author} {\bibfnamefont {P.}~\bibnamefont
  {Kunz}},\ }\href@noop {} {\enquote {\bibinfo {title} {{Finite Range Distorted
  Wave Born Approximation Code, DWUCK5}},}\ }\bibinfo {howpublished}
  {\url{http://spot.colorado.edu/~kunz/Home.html}}\BibitemShut {NoStop}%
\bibitem [{\citenamefont {Poling}\ \emph {et~al.}(1972)\citenamefont {Poling},
  \citenamefont {Norbeck},\ and\ \citenamefont {Carlson}}]{Poling_1972}%
  \BibitemOpen
  \bibfield  {author} {\bibinfo {author} {\bibfnamefont {J.~E.}\ \bibnamefont
  {Poling}}, \bibinfo {author} {\bibfnamefont {E.}~\bibnamefont {Norbeck}}, \
  and\ \bibinfo {author} {\bibfnamefont {R.~R.}\ \bibnamefont {Carlson}},\
  }\href {\doibase 10.1103/PhysRevC.5.1819} {\bibfield  {journal} {\bibinfo
  {journal} {Phys. Rev. C}\ }\textbf {\bibinfo {volume} {5}},\ \bibinfo {pages}
  {1819} (\bibinfo {year} {1972})}\BibitemShut {NoStop}%
\bibitem [{\citenamefont {Azuma}\ \emph {et~al.}(2010)\citenamefont {Azuma},
  \citenamefont {Uberseder}, \citenamefont {Simpson}, \citenamefont {Brune},
  \citenamefont {Costantini}, \citenamefont {de~Boer}, \citenamefont
  {G\"orres}, \citenamefont {Heil}, \citenamefont {LeBlanc}, \citenamefont
  {Ugalde},\ and\ \citenamefont {Wiescher}}]{AZURE2}%
  \BibitemOpen
  \bibfield  {author} {\bibinfo {author} {\bibfnamefont {R.~E.}\ \bibnamefont
  {Azuma}}, \bibinfo {author} {\bibfnamefont {E.}~\bibnamefont {Uberseder}},
  \bibinfo {author} {\bibfnamefont {E.~C.}\ \bibnamefont {Simpson}}, \bibinfo
  {author} {\bibfnamefont {C.~R.}\ \bibnamefont {Brune}}, \bibinfo {author}
  {\bibfnamefont {H.}~\bibnamefont {Costantini}}, \bibinfo {author}
  {\bibfnamefont {R.~J.}\ \bibnamefont {de~Boer}}, \bibinfo {author}
  {\bibfnamefont {J.}~\bibnamefont {G\"orres}}, \bibinfo {author}
  {\bibfnamefont {M.}~\bibnamefont {Heil}}, \bibinfo {author} {\bibfnamefont
  {P.~J.}\ \bibnamefont {LeBlanc}}, \bibinfo {author} {\bibfnamefont
  {C.}~\bibnamefont {Ugalde}}, \ and\ \bibinfo {author} {\bibfnamefont
  {M.}~\bibnamefont {Wiescher}},\ }\href {\doibase 10.1103/PhysRevC.81.045805}
  {\bibfield  {journal} {\bibinfo  {journal} {Phys. Rev. C}\ }\textbf {\bibinfo
  {volume} {81}},\ \bibinfo {pages} {045805} (\bibinfo {year}
  {2010})}\BibitemShut {NoStop}%
\bibitem [{\citenamefont {Brune}(2002)}]{Brune_2002}%
  \BibitemOpen
  \bibfield  {author} {\bibinfo {author} {\bibfnamefont {C.~R.}\ \bibnamefont
  {Brune}},\ }\href {\doibase 10.1103/PhysRevC.66.044611} {\bibfield  {journal}
  {\bibinfo  {journal} {Phys. Rev. C}\ }\textbf {\bibinfo {volume} {66}},\
  \bibinfo {pages} {044611} (\bibinfo {year} {2002})}\BibitemShut {NoStop}%
\bibitem [{ENS()}]{ENSDF}%
  \BibitemOpen
  \href@noop {} {}\bibinfo {note}
  {\url{https://www.nndc.bnl.gov/ensdf}}\BibitemShut {NoStop}%
\bibitem [{\citenamefont {Zhang}\ \emph {et~al.}(2018)\citenamefont {Zhang},
  \citenamefont {Nollett},\ and\ \citenamefont {Phillips}}]{Zhang:2017}%
  \BibitemOpen
  \bibfield  {author} {\bibinfo {author} {\bibfnamefont {X.}~\bibnamefont
  {Zhang}}, \bibinfo {author} {\bibfnamefont {K.~M.}\ \bibnamefont {Nollett}},
  \ and\ \bibinfo {author} {\bibfnamefont {D.~R.}\ \bibnamefont {Phillips}},\
  }\href {\doibase 10.1103/PhysRevC.98.034616} {\bibfield  {journal} {\bibinfo
  {journal} {Phys. Rev. C}\ }\textbf {\bibinfo {volume} {98}},\ \bibinfo
  {pages} {034616} (\bibinfo {year} {2018})}\BibitemShut {NoStop}%
\bibitem [{\citenamefont {Nollett}\ and\ \citenamefont
  {Wiringa}(2011)}]{Nollet_Wiringa}%
  \BibitemOpen
  \bibfield  {author} {\bibinfo {author} {\bibfnamefont {K.~M.}\ \bibnamefont
  {Nollett}}\ and\ \bibinfo {author} {\bibfnamefont {R.~B.}\ \bibnamefont
  {Wiringa}},\ }\href {\doibase 10.1103/PhysRevC.83.041001} {\bibfield
  {journal} {\bibinfo  {journal} {Phys. Rev. C}\ }\textbf {\bibinfo {volume}
  {83}},\ \bibinfo {pages} {041001} (\bibinfo {year} {2011})}\BibitemShut
  {NoStop}%
\bibitem [{\citenamefont {Halderson}(2006)}]{Halderson}%
  \BibitemOpen
  \bibfield  {author} {\bibinfo {author} {\bibfnamefont {D.}~\bibnamefont
  {Halderson}},\ }\href {\doibase 10.1103/PhysRevC.73.024612} {\bibfield
  {journal} {\bibinfo  {journal} {Phys. Rev. C}\ }\textbf {\bibinfo {volume}
  {73}},\ \bibinfo {pages} {024612} (\bibinfo {year} {2006})}\BibitemShut
  {NoStop}%
\bibitem [{\citenamefont {Bethe}(1949)}]{Bethe1949}%
  \BibitemOpen
  \bibfield  {author} {\bibinfo {author} {\bibfnamefont {H.~A.}\ \bibnamefont
  {Bethe}},\ }\href {\doibase 10.1103/PhysRev.76.38} {\bibfield  {journal}
  {\bibinfo  {journal} {Phys. Rev.}\ }\textbf {\bibinfo {volume} {76}},\
  \bibinfo {pages} {38} (\bibinfo {year} {1949})}\BibitemShut {NoStop}%
\bibitem [{\citenamefont {Humblet}(1985)}]{Humblet_1985}%
  \BibitemOpen
  \bibfield  {author} {\bibinfo {author} {\bibfnamefont {J.}~\bibnamefont
  {Humblet}},\ }\href {\doibase 10.1063/1.526602} {\bibfield  {journal}
  {\bibinfo  {journal} {Journal of Mathematical Physics}\ }\textbf {\bibinfo
  {volume} {26}},\ \bibinfo {pages} {656} (\bibinfo {year} {1985})},\ \Eprint
  {http://arxiv.org/abs/http://dx.doi.org/10.1063/1.526602}
  {http://dx.doi.org/10.1063/1.526602} \BibitemShut {NoStop}%
\end{thebibliography}%

\end{document}